\hsize=31pc
\vsize=49pc
\lineskip=0pt
\parskip=0pt plus 1pt
\hfuzz=1pt  
\vfuzz=2pt
\pretolerance=2500
\tolerance=5000
\vbadness=5000
\hbadness=5000
\widowpenalty=500
\clubpenalty=200
\brokenpenalty=500
\predisplaypenalty=200
\voffset=-1pc
\nopagenumbers     
\catcode`@=11
\newif\ifams
\amstrue
%
%
\amsfalse
%
%
\newfam\bdifam
\newfam\bsyfam
\newfam\bssfam
\newfam\msafam
\newfam\msbfam
\newif\ifxxpt   
\newif\ifxviipt 
\newif\ifxivpt  
\newif\ifxiipt  
\newif\ifxipt   
\newif\ifxpt    
\newif\ifixpt   
\newif\ifviiipt 
\newif\ifviipt  
\newif\ifvipt   
\newif\ifvpt    
%
%
\def\headsize#1#2{\def\headb@seline{#2}%
                \ifnum#1=20\def\HEAD{twenty}%
                           \def\smHEAD{twelve}%
                           \def\vsHEAD{nine}%
                           \ifxxpt\else\xdef\f@ntsize{\HEAD}%
                           \def\m@g{4}\def\s@ze{20.74}%
                           \loadheadfonts\xxpttrue\fi
                           \ifxiipt\else\xdef\f@ntsize{\smHEAD}%
                           \def\m@g{1}\def\s@ze{12}%
                           \loadxiiptfonts\xiipttrue\fi
                           \ifixpt\else\xdef\f@ntsize{\vsHEAD}%
                           \def\s@ze{9}%
                           \loadsmallfonts\ixpttrue\fi
                      \else
                \ifnum#1=17\def\HEAD{seventeen}%
                           \def\smHEAD{eleven}%
                           \def\vsHEAD{eight}%
                           \ifxviipt\else\xdef\f@ntsize{\HEAD}%
                           \def\m@g{3}\def\s@ze{17.28}%
                           \loadheadfonts\xviipttrue\fi
                           \ifxipt\else\xdef\f@ntsize{\smHEAD}%
                           \loadxiptfonts\xipttrue\fi
                           \ifviiipt\else\xdef\f@ntsize{\vsHEAD}%
                           \def\s@ze{8}%
                           \loadsmallfonts\viiipttrue\fi
                      \else\def\HEAD{fourteen}%
                           \def\smHEAD{ten}%
                           \def\vsHEAD{seven}%
                           \ifxivpt\else\xdef\f@ntsize{\HEAD}%
                           \def\m@g{2}\def\s@ze{14.4}%
                           \loadheadfonts\xivpttrue\fi
                           \ifxpt\else\xdef\f@ntsize{\smHEAD}%
                           \def\s@ze{10}%
                           \loadxptfonts\xpttrue\fi
                           \ifviipt\else\xdef\f@ntsize{\vsHEAD}%
                           \def\s@ze{7}%
                           \loadviiptfonts\viipttrue\fi
                \ifnum#1=14\else
                \message{Header size should be 20, 17 or 14 point
                              will now default to 14pt}\fi
                \fi\fi\headfonts}
%
%
\def\textsize#1#2{\def\textb@seline{#2}%
                 \ifnum#1=12\def\TEXT{twelve}%
                           \def\smTEXT{eight}%
                           \def\vsTEXT{six}%
                           \ifxiipt\else\xdef\f@ntsize{\TEXT}%
                           \def\m@g{1}\def\s@ze{12}%
                           \loadxiiptfonts\xiipttrue\fi
                           \ifviiipt\else\xdef\f@ntsize{\smTEXT}%
                           \def\s@ze{8}%
                           \loadsmallfonts\viiipttrue\fi
                           \ifvipt\else\xdef\f@ntsize{\vsTEXT}%
                           \def\s@ze{6}%
                           \loadviptfonts\vipttrue\fi
                      \else
                \ifnum#1=11\def\TEXT{eleven}%
                           \def\smTEXT{seven}%
                           \def\vsTEXT{five}%
                           \ifxipt\else\xdef\f@ntsize{\TEXT}%
                           \def\s@ze{11}%
                           \loadxiptfonts\xipttrue\fi
                           \ifviipt\else\xdef\f@ntsize{\smTEXT}%
                           \loadviiptfonts\viipttrue\fi
                           \ifvpt\else\xdef\f@ntsize{\vsTEXT}%
                           \def\s@ze{5}%
                           \loadvptfonts\vpttrue\fi
                      \else\def\TEXT{ten}%
                           \def\smTEXT{seven}%
                           \def\vsTEXT{five}%
                           \ifxpt\else\xdef\f@ntsize{\TEXT}%
                           \loadxptfonts\xpttrue\fi
                           \ifviipt\else\xdef\f@ntsize{\smTEXT}%
                           \def\s@ze{7}%
                           \loadviiptfonts\viipttrue\fi
                           \ifvpt\else\xdef\f@ntsize{\vsTEXT}%
                           \def\s@ze{5}%
                           \loadvptfonts\vpttrue\fi
                \ifnum#1=10\else
                \message{Text size should be 12, 11 or 10 point
                              will now default to 10pt}\fi
                \fi\fi\textfonts}
%
%
\def\smallsize#1#2{\def\smallb@seline{#2}%
                 \ifnum#1=10\def\SMALL{ten}%
                           \def\smSMALL{seven}%
                           \def\vsSMALL{five}%
                           \ifxpt\else\xdef\f@ntsize{\SMALL}%
                           \loadxptfonts\xpttrue\fi
                           \ifviipt\else\xdef\f@ntsize{\smSMALL}%
                           \def\s@ze{7}%
                           \loadviiptfonts\viipttrue\fi
                           \ifvpt\else\xdef\f@ntsize{\vsSMALL}%
                           \def\s@ze{5}%
                           \loadvptfonts\vpttrue\fi
                       \else
                 \ifnum#1=9\def\SMALL{nine}%
                           \def\smSMALL{six}%
                           \def\vsSMALL{five}%
                           \ifixpt\else\xdef\f@ntsize{\SMALL}%
                           \def\s@ze{9}%
                           \loadsmallfonts\ixpttrue\fi
                           \ifvipt\else\xdef\f@ntsize{\smSMALL}%
                           \def\s@ze{6}%
                           \loadviptfonts\vipttrue\fi
                           \ifvpt\else\xdef\f@ntsize{\vsSMALL}%
                           \def\s@ze{5}%
                           \loadvptfonts\vpttrue\fi
                       \else
                           \def\SMALL{eight}%
                           \def\smSMALL{six}%
                           \def\vsSMALL{five}%
                           \ifviiipt\else\xdef\f@ntsize{\SMALL}%
                           \def\s@ze{8}%
                           \loadsmallfonts\viiipttrue\fi
                           \ifvipt\else\xdef\f@ntsize{\smSMALL}%
                           \def\s@ze{6}%
                           \loadviptfonts\vipttrue\fi
                           \ifvpt\else\xdef\f@ntsize{\vsSMALL}%
                           \def\s@ze{5}%
                           \loadvptfonts\vpttrue\fi
                 \ifnum#1=8\else\message{Small size should be 10, 9 or 
                            8 point will now default to 8pt}\fi
                \fi\fi\smallfonts}
\def\F@nt{\expandafter\font\csname}
\def\Sk@w{\expandafter\skewchar\csname}
\def\@nd{\endcsname}
\def\@step#1{ scaled \magstep#1}
\def\@half{ scaled \magstephalf}
\def\@t#1{ at #1pt}
%
%
\def\loadheadfonts{\bigf@nts
\F@nt \f@ntsize bdi\@nd=cmmib10 \@t{\s@ze}%
\Sk@w \f@ntsize bdi\@nd='177
\F@nt \f@ntsize bsy\@nd=cmbsy10 \@t{\s@ze}%
\Sk@w \f@ntsize bsy\@nd='60
\F@nt \f@ntsize bss\@nd=cmssbx10 \@t{\s@ze}}
%
%
\def\loadxiiptfonts{\bigf@nts
\F@nt \f@ntsize bdi\@nd=cmmib10 \@step{\m@g}%
\Sk@w \f@ntsize bdi\@nd='177
\F@nt \f@ntsize bsy\@nd=cmbsy10 \@step{\m@g}%
\Sk@w \f@ntsize bsy\@nd='60
\F@nt \f@ntsize bss\@nd=cmssbx10 \@step{\m@g}}
%
%
\def\loadxiptfonts{%
\font\elevenrm=cmr10 \@half
\font\eleveni=cmmi10 \@half
\skewchar\eleveni='177
\font\elevensy=cmsy10 \@half
\skewchar\elevensy='60
\font\elevenex=cmex10 \@half
\font\elevenit=cmti10 \@half
\font\elevensl=cmsl10 \@half
\font\elevenbf=cmbx10 \@half
\font\eleventt=cmtt10 \@half
\ifams\font\elevenmsa=msam10 \@half
\font\elevenmsb=msbm10 \@half\else\fi
\font\elevenbdi=cmmib10 \@half
\skewchar\elevenbdi='177
\font\elevenbsy=cmbsy10 \@half
\skewchar\elevenbsy='60
\font\elevenbss=cmssbx10 \@half}
%
%
\def\loadxptfonts{%
\font\tenbdi=cmmib10
\skewchar\tenbdi='177
\font\tenbsy=cmbsy10 
\skewchar\tenbsy='60
\ifams\font\tenmsa=msam10 
\font\tenmsb=msbm10\else\fi
\font\tenbss=cmssbx10}%
%
%
\def\loadsmallfonts{\smallf@nts
\ifams
\F@nt \f@ntsize ex\@nd=cmex\s@ze
\else
\F@nt \f@ntsize ex\@nd=cmex10\fi
\F@nt \f@ntsize it\@nd=cmti\s@ze
\F@nt \f@ntsize sl\@nd=cmsl\s@ze
\F@nt \f@ntsize tt\@nd=cmtt\s@ze}
%
%
\def\loadviiptfonts{%
\font\sevenit=cmti7
\font\sevensl=cmsl8 at 7pt
\ifams\font\sevenmsa=msam7 
\font\sevenmsb=msbm7
\font\sevenex=cmex7
\font\sevenbsy=cmbsy7
\font\sevenbdi=cmmib7\else
\font\sevenex=cmex10
\font\sevenbsy=cmbsy10 at 7pt
\font\sevenbdi=cmmib10 at 7pt\fi
\skewchar\sevenbsy='60
\skewchar\sevenbdi='177
\font\sevenbss=cmssbx10 at 7pt}%
%
%
\def\loadviptfonts{\smallf@nts
\ifams\font\sixex=cmex7 at 6pt\else
\font\sixex=cmex10\fi
\font\sixit=cmti7 at 6pt}
%
%
\def\loadvptfonts{%
\font\fiveit=cmti7 at 5pt
\ifams\font\fiveex=cmex7 at 5pt
\font\fivebdi=cmmib5
\font\fivebsy=cmbsy5
\font\fivemsa=msam5 
\font\fivemsb=msbm5\else
\font\fiveex=cmex10
\font\fivebdi=cmmib10 at 5pt
\font\fivebsy=cmbsy10 at 5pt\fi
\skewchar\fivebdi='177
\skewchar\fivebsy='60
\font\fivebss=cmssbx10 at 5pt}
\def\bigf@nts{%
\F@nt \f@ntsize rm\@nd=cmr10 \@step{\m@g}%
\F@nt \f@ntsize i\@nd=cmmi10 \@step{\m@g}%
\Sk@w \f@ntsize i\@nd='177
\F@nt \f@ntsize sy\@nd=cmsy10 \@step{\m@g}%
\Sk@w \f@ntsize sy\@nd='60
\F@nt \f@ntsize ex\@nd=cmex10 \@step{\m@g}%
\F@nt \f@ntsize it\@nd=cmti10 \@step{\m@g}%
\F@nt \f@ntsize sl\@nd=cmsl10 \@step{\m@g}%
\F@nt \f@ntsize bf\@nd=cmbx10 \@step{\m@g}%
\F@nt \f@ntsize tt\@nd=cmtt10 \@step{\m@g}%
\ifams
\F@nt \f@ntsize msa\@nd=msam10 \@step{\m@g}%
\F@nt \f@ntsize msb\@nd=msbm10 \@step{\m@g}\else\fi}
\def\smallf@nts{%
\F@nt \f@ntsize rm\@nd=cmr\s@ze
\F@nt \f@ntsize i\@nd=cmmi\s@ze 
\Sk@w \f@ntsize i\@nd='177
\F@nt \f@ntsize sy\@nd=cmsy\s@ze
\Sk@w \f@ntsize sy\@nd='60
\F@nt \f@ntsize bf\@nd=cmbx\s@ze 
\ifams
\F@nt \f@ntsize bdi\@nd=cmmib\s@ze 
\F@nt \f@ntsize bsy\@nd=cmbsy\s@ze 
\F@nt \f@ntsize msa\@nd=msam\s@ze 
\F@nt \f@ntsize msb\@nd=msbm\s@ze
\else
\F@nt \f@ntsize bdi\@nd=cmmib10 \@t{\s@ze}%
\F@nt \f@ntsize bsy\@nd=cmbsy10 \@t{\s@ze}\fi 
\Sk@w \f@ntsize bdi\@nd='177
\Sk@w \f@ntsize bsy\@nd='60
\F@nt \f@ntsize bss\@nd=cmssbx10 \@t{\s@ze}}%
%
%
\def\headfonts{%
\textfont0=\csname\HEAD rm\@nd        
\scriptfont0=\csname\smHEAD rm\@nd
\scriptscriptfont0=\csname\vsHEAD rm\@nd
\def\rm{\fam0\csname\HEAD rm\@nd
\def\sc{\csname\smHEAD rm\@nd}}%
\textfont1=\csname\HEAD i\@nd         
\scriptfont1=\csname\smHEAD i\@nd
\scriptscriptfont1=\csname\vsHEAD i\@nd
\textfont2=\csname\HEAD sy\@nd        
\scriptfont2=\csname\smHEAD sy\@nd
\scriptscriptfont2=\csname\vsHEAD sy\@nd
\textfont3=\csname\HEAD ex\@nd        
\scriptfont3=\csname\smHEAD ex\@nd
\scriptscriptfont3=\csname\smHEAD ex\@nd
\textfont\itfam=\csname\HEAD it\@nd   
\scriptfont\itfam=\csname\smHEAD it\@nd
\scriptscriptfont\itfam=\csname\vsHEAD it\@nd
\def\it{\fam\itfam\csname\HEAD it\@nd
\def\sc{\csname\smHEAD it\@nd}}%
\textfont\slfam=\csname\HEAD sl\@nd   
\def\sl{\fam\slfam\csname\HEAD sl\@nd
\def\sc{\csname\smHEAD sl\@nd}}%
\textfont\bffam=\csname\HEAD bf\@nd   
\scriptfont\bffam=\csname\smHEAD bf\@nd
\scriptscriptfont\bffam=\csname\vsHEAD bf\@nd
\def\bf{\fam\bffam\csname\HEAD bf\@nd
\def\sc{\csname\smHEAD bf\@nd}}%
\textfont\ttfam=\csname\HEAD tt\@nd   
\def\tt{\fam\ttfam\csname\HEAD tt\@nd}%
\textfont\bdifam=\csname\HEAD bdi\@nd 
\scriptfont\bdifam=\csname\smHEAD bdi\@nd
\scriptscriptfont\bdifam=\csname\vsHEAD bdi\@nd
\def\bdi{\fam\bdifam\csname\HEAD bdi\@nd}%
\textfont\bsyfam=\csname\HEAD bsy\@nd 
\scriptfont\bsyfam=\csname\smHEAD bsy\@nd
\def\bsy{\fam\bsyfam\csname\HEAD bsy\@nd}%
\textfont\bssfam=\csname\HEAD bss\@nd 
\scriptfont\bssfam=\csname\smHEAD bss\@nd
\scriptscriptfont\bssfam=\csname\vsHEAD bss\@nd
\def\bss{\fam\bssfam\csname\HEAD bss\@nd}%
\ifams
\textfont\msafam=\csname\HEAD msa\@nd 
\scriptfont\msafam=\csname\smHEAD msa\@nd
\scriptscriptfont\msafam=\csname\vsHEAD msa\@nd
\textfont\msbfam=\csname\HEAD msb\@nd 
\scriptfont\msbfam=\csname\smHEAD msb\@nd
\scriptscriptfont\msbfam=\csname\vsHEAD msb\@nd
\else\fi
\normalbaselineskip=\headb@seline pt%
\setbox\strutbox=\hbox{\vrule height.7\normalbaselineskip 
depth.3\baselineskip width0pt}%
\def\sc{\csname\smHEAD rm\@nd}\normalbaselines\bf}
%
%
\def\textfonts{%
\textfont0=\csname\TEXT rm\@nd        
\scriptfont0=\csname\smTEXT rm\@nd
\scriptscriptfont0=\csname\vsTEXT rm\@nd
\def\rm{\fam0\csname\TEXT rm\@nd
\def\sc{\csname\smTEXT rm\@nd}}%
\textfont1=\csname\TEXT i\@nd         
\scriptfont1=\csname\smTEXT i\@nd
\scriptscriptfont1=\csname\vsTEXT i\@nd
\textfont2=\csname\TEXT sy\@nd        
\scriptfont2=\csname\smTEXT sy\@nd
\scriptscriptfont2=\csname\vsTEXT sy\@nd
\textfont3=\csname\TEXT ex\@nd        
\scriptfont3=\csname\smTEXT ex\@nd
\scriptscriptfont3=\csname\smTEXT ex\@nd
\textfont\itfam=\csname\TEXT it\@nd   
\scriptfont\itfam=\csname\smTEXT it\@nd
\scriptscriptfont\itfam=\csname\vsTEXT it\@nd
\def\it{\fam\itfam\csname\TEXT it\@nd
\def\sc{\csname\smTEXT it\@nd}}%
\textfont\slfam=\csname\TEXT sl\@nd   
\def\sl{\fam\slfam\csname\TEXT sl\@nd
\def\sc{\csname\smTEXT sl\@nd}}%
\textfont\bffam=\csname\TEXT bf\@nd   
\scriptfont\bffam=\csname\smTEXT bf\@nd
\scriptscriptfont\bffam=\csname\vsTEXT bf\@nd
\def\bf{\fam\bffam\csname\TEXT bf\@nd
\def\sc{\csname\smTEXT bf\@nd}}%
\textfont\ttfam=\csname\TEXT tt\@nd   
\def\tt{\fam\ttfam\csname\TEXT tt\@nd}%
\textfont\bdifam=\csname\TEXT bdi\@nd 
\scriptfont\bdifam=\csname\smTEXT bdi\@nd
\scriptscriptfont\bdifam=\csname\vsTEXT bdi\@nd
\def\bdi{\fam\bdifam\csname\TEXT bdi\@nd}%
\textfont\bsyfam=\csname\TEXT bsy\@nd 
\scriptfont\bsyfam=\csname\smTEXT bsy\@nd
\def\bsy{\fam\bsyfam\csname\TEXT bsy\@nd}%
\textfont\bssfam=\csname\TEXT bss\@nd 
\scriptfont\bssfam=\csname\smTEXT bss\@nd
\scriptscriptfont\bssfam=\csname\vsTEXT bss\@nd
\def\bss{\fam\bssfam\csname\TEXT bss\@nd}%
\ifams
\textfont\msafam=\csname\TEXT msa\@nd 
\scriptfont\msafam=\csname\smTEXT msa\@nd
\scriptscriptfont\msafam=\csname\vsTEXT msa\@nd
\textfont\msbfam=\csname\TEXT msb\@nd 
\scriptfont\msbfam=\csname\smTEXT msb\@nd
\scriptscriptfont\msbfam=\csname\vsTEXT msb\@nd
\else\fi
\normalbaselineskip=\textb@seline pt
\setbox\strutbox=\hbox{\vrule height.7\normalbaselineskip 
depth.3\baselineskip width0pt}%
\everymath{}%
\def\sc{\csname\smTEXT rm\@nd}\normalbaselines\rm}
%
%
\def\smallfonts{%
\textfont0=\csname\SMALL rm\@nd        
\scriptfont0=\csname\smSMALL rm\@nd
\scriptscriptfont0=\csname\vsSMALL rm\@nd
\def\rm{\fam0\csname\SMALL rm\@nd
\def\sc{\csname\smSMALL rm\@nd}}%
\textfont1=\csname\SMALL i\@nd         
\scriptfont1=\csname\smSMALL i\@nd
\scriptscriptfont1=\csname\vsSMALL i\@nd
\textfont2=\csname\SMALL sy\@nd        
\scriptfont2=\csname\smSMALL sy\@nd
\scriptscriptfont2=\csname\vsSMALL sy\@nd
\textfont3=\csname\SMALL ex\@nd        
\scriptfont3=\csname\smSMALL ex\@nd
\scriptscriptfont3=\csname\smSMALL ex\@nd
\textfont\itfam=\csname\SMALL it\@nd   
\scriptfont\itfam=\csname\smSMALL it\@nd
\scriptscriptfont\itfam=\csname\vsSMALL it\@nd
\def\it{\fam\itfam\csname\SMALL it\@nd
\def\sc{\csname\smSMALL it\@nd}}%
\textfont\slfam=\csname\SMALL sl\@nd   
\def\sl{\fam\slfam\csname\SMALL sl\@nd
\def\sc{\csname\smSMALL sl\@nd}}%
\textfont\bffam=\csname\SMALL bf\@nd   
\scriptfont\bffam=\csname\smSMALL bf\@nd
\scriptscriptfont\bffam=\csname\vsSMALL bf\@nd
\def\bf{\fam\bffam\csname\SMALL bf\@nd
\def\sc{\csname\smSMALL bf\@nd}}%
\textfont\ttfam=\csname\SMALL tt\@nd   
\def\tt{\fam\ttfam\csname\SMALL tt\@nd}%
\textfont\bdifam=\csname\SMALL bdi\@nd 
\scriptfont\bdifam=\csname\smSMALL bdi\@nd
\scriptscriptfont\bdifam=\csname\vsSMALL bdi\@nd
\def\bdi{\fam\bdifam\csname\SMALL bdi\@nd}%
\textfont\bsyfam=\csname\SMALL bsy\@nd 
\scriptfont\bsyfam=\csname\smSMALL bsy\@nd
\def\bsy{\fam\bsyfam\csname\SMALL bsy\@nd}%
\textfont\bssfam=\csname\SMALL bss\@nd 
\scriptfont\bssfam=\csname\smSMALL bss\@nd
\scriptscriptfont\bssfam=\csname\vsSMALL bss\@nd
\def\bss{\fam\bssfam\csname\SMALL bss\@nd}%
\ifams
\textfont\msafam=\csname\SMALL msa\@nd 
\scriptfont\msafam=\csname\smSMALL msa\@nd
\scriptscriptfont\msafam=\csname\vsSMALL msa\@nd
\textfont\msbfam=\csname\SMALL msb\@nd 
\scriptfont\msbfam=\csname\smSMALL msb\@nd
\scriptscriptfont\msbfam=\csname\vsSMALL msb\@nd
\else\fi
\normalbaselineskip=\smallb@seline pt%
\setbox\strutbox=\hbox{\vrule height.7\normalbaselineskip 
depth.3\baselineskip width0pt}%
\everymath{}%
\def\sc{\csname\smSMALL rm\@nd}\normalbaselines\rm}%
\everydisplay{\indenteddisplay
   \gdef\labeltype{\eqlabel}}%
%
%
\def\hexnumber@#1{\ifcase#1 0\or 1\or 2\or 3\or 4\or 5\or 6\or 7\or 8\or
 9\or A\or B\or C\or D\or E\or F\fi}
\edef\bffam@{\hexnumber@\bffam}
\edef\bdifam@{\hexnumber@\bdifam}
\edef\bsyfam@{\hexnumber@\bsyfam}
\def\undefine#1{\let#1\undefined}
\def\newsymbol#1#2#3#4#5{\let\next@\relax
 \ifnum#2=\thr@@\let\next@\bdifam@\else
 \ifams
 \ifnum#2=\@ne\let\next@\msafam@\else
 \ifnum#2=\tw@\let\next@\msbfam@\fi\fi
 \fi\fi
 \mathchardef#1="#3\next@#4#5}
\def\mathhexbox@#1#2#3{\relax
 \ifmmode\mathpalette{}{\m@th\mathchar"#1#2#3}%
 \else\leavevmode\hbox{$\m@th\mathchar"#1#2#3$}\fi}

\def\bi#1{{\fam\bdifam\relax#1}}
%
%
\ifams\input amsmacro\fi
%
%
\newsymbol\bitGamma 3000
\newsymbol\bitDelta 3001
\newsymbol\bitTheta 3002
\newsymbol\bitLambda 3003
\newsymbol\bitXi 3004
\newsymbol\bitPi 3005
\newsymbol\bitSigma 3006
\newsymbol\bitUpsilon 3007
\newsymbol\bitPhi 3008
\newsymbol\bitPsi 3009
\newsymbol\bitOmega 300A
\newsymbol\balpha 300B
\newsymbol\bbeta 300C
\newsymbol\bgamma 300D
\newsymbol\bdelta 300E
\newsymbol\bepsilon 300F
\newsymbol\bzeta 3010
\newsymbol\bfeta 3011
\newsymbol\btheta 3012
\newsymbol\biota 3013
\newsymbol\bkappa 3014
\newsymbol\blambda 3015
\newsymbol\bmu 3016
\newsymbol\bnu 3017
\newsymbol\bxi 3018
\newsymbol\bpi 3019
\newsymbol\brho 301A
\newsymbol\bsigma 301B
\newsymbol\btau 301C
\newsymbol\bupsilon 301D
\newsymbol\bphi 301E
\newsymbol\bchi 301F
\newsymbol\bpsi 3020
\newsymbol\bomega 3021
\newsymbol\bvarepsilon 3022
\newsymbol\bvartheta 3023
\newsymbol\bvaromega 3024
\newsymbol\bvarrho 3025
\newsymbol\bvarzeta 3026
\newsymbol\bvarphi 3027
\newsymbol\bpartial 3040
\newsymbol\bell 3060
\newsymbol\bimath 307B
\newsymbol\bjmath 307C
\mathchardef\binfty "0\bsyfam@31
\mathchardef\bnabla "0\bsyfam@72
\mathchardef\bdot "2\bsyfam@01
\mathchardef\bGamma "0\bffam@00
\mathchardef\bDelta "0\bffam@01
\mathchardef\bTheta "0\bffam@02
\mathchardef\bLambda "0\bffam@03
\mathchardef\bXi "0\bffam@04
\mathchardef\bPi "0\bffam@05
\mathchardef\bSigma "0\bffam@06
\mathchardef\bUpsilon "0\bffam@07
\mathchardef\bPhi "0\bffam@08
\mathchardef\bPsi "0\bffam@09
\mathchardef\bOmega "0\bffam@0A
\mathchardef\itGamma "0100
\mathchardef\itDelta "0101
\mathchardef\itTheta "0102
\mathchardef\itLambda "0103
\mathchardef\itXi "0104
\mathchardef\itPi "0105
\mathchardef\itSigma "0106
\mathchardef\itUpsilon "0107
\mathchardef\itPhi "0108
\mathchardef\itPsi "0109
\mathchardef\itOmega "010A
\mathchardef\Gamma "0000
\mathchardef\Delta "0001
\mathchardef\Theta "0002
\mathchardef\Lambda "0003
\mathchardef\Xi "0004
\mathchardef\Pi "0005
\mathchardef\Sigma "0006
\mathchardef\Upsilon "0007
\mathchardef\Phi "0008
\mathchardef\Psi "0009
\mathchardef\Omega "000A
%
%
\newcount\firstpage  \firstpage=1  
\newcount\jnl                      
\newcount\secno                    
\newcount\subno                    
\newcount\subsubno                 
\newcount\appno                    
\newcount\tabno                    
\newcount\figno                    
\newcount\countno                  
\newcount\refno                    
\newcount\eqlett     \eqlett=97    
\newif\ifletter
\newif\ifwide
\newif\ifnotfull
\newif\ifaligned
\newif\ifnumbysec  
\newif\ifappendix
\newif\ifnumapp
\newif\ifssf
\newif\ifppt
\newdimen\t@bwidth
\newdimen\c@pwidth
\newdimen\digitwidth                    
\newdimen\argwidth                      
\newdimen\secindent    \secindent=5pc   
\newdimen\textind    \textind=16pt      
\newdimen\tempval                       
\newskip\beforesecskip
\def\beforesecspace{\vskip\beforesecskip\relax}
\newskip\beforesubskip
\def\beforesubspace{\vskip\beforesubskip\relax}
\newskip\beforesubsubskip
\def\beforesubsubspace{\vskip\beforesubsubskip\relax}
\newskip\secskip
\def\secspace{\vskip\secskip\relax}
\newskip\subskip
\def\subspace{\vskip\subskip\relax}
\newskip\insertskip
\def\insertspace{\vskip\insertskip\relax}
\def\sp@ce{\ifx\next*\let\next=\@ssf
               \else\let\next=\@nossf\fi\next}
\def\@ssf#1{\nobreak\secspace\global\ssftrue\nobreak}
\def\@nossf{\nobreak\secspace\nobreak\noindent\ignorespaces}
\def\subsp@ce{\ifx\next*\let\next=\@sssf
               \else\let\next=\@nosssf\fi\next}
\def\@sssf#1{\nobreak\subspace\global\ssftrue\nobreak}
\def\@nosssf{\nobreak\subspace\nobreak\noindent\ignorespaces}
\beforesecskip=24pt plus12pt minus8pt
\beforesubskip=12pt plus6pt minus4pt
\beforesubsubskip=12pt plus6pt minus4pt
\secskip=12pt plus 2pt minus 2pt
\subskip=6pt plus3pt minus2pt
\insertskip=18pt plus6pt minus6pt%
\fontdimen16\tensy=2.7pt
\fontdimen17\tensy=2.7pt
%
%
\def\eqlabel{(\ifappendix\applett
               \ifnumbysec\ifnum\secno>0 \the\secno\fi.\fi
               \else\ifnumbysec\the\secno.\fi\fi\the\countno)}
\def\seclabel{\ifappendix\ifnumapp\else\applett\fi
    \ifnum\secno>0 \the\secno
    \ifnumbysec\ifnum\subno>0.\the\subno\fi\fi\fi
    \else\the\secno\fi\ifnum\subno>0.\the\subno
         \ifnum\subsubno>0.\the\subsubno\fi\fi}
\def\tablabel{\ifappendix\applett\fi\the\tabno}
\def\figlabel{\ifappendix\applett\fi\the\figno}
\def\gac{\global\advance\countno by 1}
%
%

\def\vfootnote#1{\insert\footins\bgroup
\interlinepenalty=\interfootnotelinepenalty
\splittopskip=\ht\strutbox 
\splitmaxdepth=\dp\strutbox \floatingpenalty=20000
\leftskip=0pt \rightskip=0pt \spaceskip=0pt \xspaceskip=0pt%
\noindent\smallfonts\rm #1\ \ignorespaces\footstrut\futurelet\next\fo@t}
%
%
\def\endinsert{\egroup
    \if@mid \dimen@=\ht0 \advance\dimen@ by\dp0
       \advance\dimen@ by12\p@ \advance\dimen@ by\pagetotal
       \ifdim\dimen@>\pagegoal \@midfalse\p@gefalse\fi\fi
    \if@mid \insertspace \box0 \par \ifdim\lastskip<\insertskip
    \removelastskip \penalty-200 \insertspace \fi
    \else\insert\topins{\penalty100
       \splittopskip=0pt \splitmaxdepth=\maxdimen 
       \floatingpenalty=0
       \ifp@ge \dimen@=\dp0
       \vbox to\vsize{\unvbox0 \kern-\dimen@}%
       \else\box0\nobreak\insertspace\fi}\fi\endgroup}   
%
%
%
\def\ind{\hbox to \secindent{\hfill}}
%
%

%
%
\def\lo#1{\llap{${}#1{}$}}
%
%
\def\indeqn#1{\alignedfalse\displ@y\halign{\hbox to \displaywidth
    {$\ind\@lign\displaystyle##\hfil$}\crcr #1\crcr}}
%
%
\def\indalign#1{\alignedtrue\displ@y \tabskip=0pt 
  \halign to\displaywidth{\ind$\@lign\displaystyle{##}$\tabskip=0pt
    &$\@lign\displaystyle{{}##}$\hfill\tabskip=\centering
    &\llap{$\@lign\hbox{\rm##}$}\tabskip=0pt\crcr
    #1\crcr}}
\def\fl{{\hskip-\secindent}}
\def\indenteddisplay#1$${\indispl@y{#1 }}
\def\indispl@y#1{\disptest#1\eqalignno\eqalignno\disptest}
\def\disptest#1\eqalignno#2\eqalignno#3\disptest{%
    \ifx#3\eqalignno
    \indalign#2%
    \else\indeqn{#1}\fi$$}
%
%

%
%

%
%

%
%

%
%

\def\ns{\noalign{\vskip-3pt}}

%

%
%
\def\bhbar{\rlap{\kern1pt\raise.4ex\hbox{\bf\char'40}}\bi{h}}

\def\frac#1#2{{#1\over#2}}
\ifams
\def\lap{\lesssim}
\def\gap{\gtrsim}

\let\geq=\geqslant
\else

\def\gap{\;\lower3pt\hbox{$\buildrel > \over \sim$}\;}%
\def\lap{\;\lower3pt\hbox{$\buildrel < \over \sim$}\;}\fi

\chardef\ii="10
\def\tqs{\hbox to 25pt{\hfil}}

\def\Bbbone{1\kern-.22em {\rm l}}
%
%
\def\rp{\raise8pt\hbox{$\scriptstyle\prime$}}
%
%
%
%

%
%
\def\[#1\]{\setbox0=\hbox{$\dsty#1$}\argwidth=\wd0
    \setbox0=\hbox{$\left[\box0\right]$}\advance\argwidth by -\wd0
    \left[\kern.3\argwidth\box0\kern.3\argwidth\right]}
%
%
\def\lsb#1\rsb{\setbox0=\hbox{$#1$}\argwidth=\wd0
    \setbox0=\hbox{$\left[\box0\right]$}\advance\argwidth by -\wd0
    \left[\kern.3\argwidth\box0\kern.3\argwidth\right]}
%

%
%

%
\def\pt(#1){({\it #1\/})}
\let\dsty=\displaystyle

%
%
\def\reactions#1{\vskip 12pt plus2pt minus2pt%
\vbox{\hbox{\kern\secindent\vrule\kern12pt%
\vbox{\kern0.5pt\vbox{\hsize=24pc\parindent=0pt\smallfonts\rm NUCLEAR 
REACTIONS\strut\quad #1\strut}\kern0.5pt}\kern12pt\vrule}}}
%
%
\def\slashchar#1{\setbox0=\hbox{$#1$}\dimen0=\wd0%
\setbox1=\hbox{/}\dimen1=\wd1%
\ifdim\dimen0>\dimen1%
\rlap{\hbox to \dimen0{\hfil/\hfil}}#1\else                                        
\rlap{\hbox to \dimen1{\hfil$#1$\hfil}}/\fi}
%
%
\def\textindent#1{\noindent\hbox to \parindent{#1\hss}\ignorespaces}
%
%
\def\opencirc{\raise1pt\hbox{$\scriptstyle{\bigcirc}$}}

\ifams
\def\opensqr{\hbox{$\square$}}

\def\opentridown{\hbox{$\triangledown$}}

\else
\def\opensqr{\vbox{\hrule height.4pt\hbox{\vrule width.4pt height3.5pt
    \kern3.5pt\vrule width.4pt}\hrule height.4pt}}

\def\opentridown{\raise1pt\hbox{$\scriptstyle\bigtriangledown$}}

\fi

%
%
\def\m@th{\mathsurround=0pt}
%
%
\def\cases#1{%
\left\{\,\vcenter{\normalbaselines\openup1\jot\m@th%
     \ialign{$\displaystyle##\hfil$&\rm\tqs##\hfil\crcr#1\crcr}}\right.}%
%
%
\def\oldcases#1{\left\{\,\vcenter{\normalbaselines\m@th
    \ialign{$##\hfil$&\rm\quad##\hfil\crcr#1\crcr}}\right.}
%
%
\def\numcases#1{\left\{\,\vcenter{\baselineskip=15pt\m@th%
     \ialign{$\displaystyle##\hfil$&\rm\tqs##\hfil
     \crcr#1\crcr}}\right.\hfill
     \vcenter{\baselineskip=15pt\m@th%
     \ialign{\rlap{$\phantom{\displaystyle##\hfil}$}\tabskip=0pt&\en
     \rlap{\phantom{##\hfil}}\crcr#1\crcr}}}
\def\ptnumcases#1{\left\{\,\vcenter{\baselineskip=15pt\m@th%
     \ialign{$\displaystyle##\hfil$&\rm\tqs##\hfil
     \crcr#1\crcr}}\right.\hfill
     \vcenter{\baselineskip=15pt\m@th%
     \ialign{\rlap{$\phantom{\displaystyle##\hfil}$}\tabskip=0pt&\enpt
     \rlap{\phantom{##\hfil}}\crcr#1\crcr}}\global\eqlett=97
     \global\advance\countno by 1}
%
%
\def\eq(#1){\ifaligned\@mp(#1)\else\hfill\llap{{\rm (#1)}}\fi}
\def\ceq(#1){\ns\ns\ifaligned\@mp\fi\eq(#1)\cr\ns\ns}
\def\eqpt(#1#2){\ifaligned\@mp(#1{\it #2\/})
                    \else\hfill\llap{{\rm (#1{\it #2\/})}}\fi}

%
%
\countno=1

\def\aleq{&\rm(\ifappendix\applett
               \ifnumbysec\ifnum\secno>0 \the\secno\fi.\fi
               \else\ifnumbysec\the\secno.\fi\fi\the\countno}
\def\noaleq{\hfill\llap\bgroup\rm(\ifappendix\applett
               \ifnumbysec\ifnum\secno>0 \the\secno\fi.\fi
               \else\ifnumbysec\the\secno.\fi\fi\the\countno}
\def\@mp{&}
\def\en{\ifaligned\aleq)\else\noaleq)\egroup\fi\gac}
\def\cen{\ns\ns\ifaligned\@mp\fi\en\cr\ns\ns}
\def\enpt{\ifaligned\aleq{\it\char\the\eqlett})\else
    \noaleq{\it\char\the\eqlett})\egroup\fi
    \global\advance\eqlett by 1}
\def\endpt{\ifaligned\aleq{\it\char\the\eqlett})\else
    \noaleq{\it\char\the\eqlett})\egroup\fi
    \global\eqlett=97\gac}
%
%




%
%

%
\headline={\ifodd\pageno{\ifnum\pageno=\firstpage\hfill
   \else\rrhead\fi}\else\lrhead\fi}
\def\rrhead{\textfonts\hskip\secindent\it
    \shorttitle\hfill\rm\folio}
\def\lrhead{\textfonts\hbox to\secindent{\rm\folio\hss}%
    \it\aunames\hss}
\footline={\ifnum\pageno=\firstpage \hfill\textfonts\rm\folio\fi}
\def\@rticle#1#2{\vglue.5pc
    {\parindent=\secindent \bf #1\par}
     \vskip2.5pc
    {\exhyphenpenalty=10000\hyphenpenalty=10000
     \baselineskip=18pt\raggedright\noindent
     \headfonts\bf#2\par}\futurelet\next\sh@rttitle}%
\def\title#1{\gdef\shorttitle{#1}
    \vglue4pc{\exhyphenpenalty=10000\hyphenpenalty=10000 
    \baselineskip=18pt 
    \raggedright\parindent=0pt
    \headfonts\bf#1\par}\futurelet\next\sh@rttitle} 

\def\article#1#2{\gdef\shorttitle{#2}\@rticle{#1}{#2}} 
\def\review#1{\gdef\shorttitle{#1}%
    \@rticle{REVIEW \ifpbm\else ARTICLE\fi}{#1}}
\def\topical#1{\gdef\shorttitle{#1}%
    \@rticle{TOPICAL REVIEW}{#1}}
\def\comment#1{\gdef\shorttitle{#1}%
    \@rticle{COMMENT}{#1}}
\def\note#1{\gdef\shorttitle{#1}%
    \@rticle{NOTE}{#1}}
\def\prelim#1{\gdef\shorttitle{#1}%
    \@rticle{PRELIMINARY COMMUNICATION}{#1}}
\def\letter#1{\gdef\shorttitle{Letter to the Editor}%
     \gdef\aunames{Letter to the Editor}
     \global\lettertrue\ifnum\jnl=7\global\letterfalse\fi
     \@rticle{LETTER TO THE EDITOR}{#1}}
\def\sh@rttitle{\ifx\next[\let\next=\sh@rt
                \else\let\next=\f@ll\fi\next}
\def\sh@rt[#1]{\gdef\shorttitle{#1}}
\def\f@ll{}
\def\author#1{\ifletter\else\gdef\aunames{#1}\fi\vskip1.5pc
    {\parindent=\secindent  
     \hang\textfonts  
     \ifppt\bf\else\rm\fi#1\par}  
     \ifppt\bigskip\else\smallskip\fi
     \futurelet\next\@unames}
\def\@unames{\ifx\next[\let\next=\short@uthor
                 \else\let\next=\@uthor\fi\next}
\def\short@uthor[#1]{\gdef\aunames{#1}}
\def\@uthor{}
\def\address#1{{\parindent=\secindent
    \exhyphenpenalty=10000\hyphenpenalty=10000
\ifppt\textfonts\else\smallfonts\fi\hang\raggedright\rm#1\par}%
    \ifppt\bigskip\fi}
\def\jl#1{\global\jnl=#1}
\jl{0}%
\def\journal{\ifnum\jnl=1 J. Phys.\ A: Math.\ Gen.\ 
        \else\ifnum\jnl=2 J. Phys.\ B: At.\ Mol.\ Opt.\ Phys.\ 
        \else\ifnum\jnl=3 J. Phys.:\ Condens. Matter\ 
        \else\ifnum\jnl=4 J. Phys.\ G: Nucl.\ Part.\ Phys.\ 
        \else\ifnum\jnl=5 Inverse Problems\ 
        \else\ifnum\jnl=6 Class. Quantum Grav.\ 
        \else\ifnum\jnl=7 Network\ 
        \else\ifnum\jnl=8 Nonlinearity\
        \else\ifnum\jnl=9 Quantum Opt.\
        \else\ifnum\jnl=10 Waves in Random Media\
        \else\ifnum\jnl=11 Pure Appl. Opt.\ 
        \else\ifnum\jnl=12 Phys. Med. Biol.\
        \else\ifnum\jnl=13 Modelling Simulation Mater.\ Sci.\ Eng.\ 
        \else\ifnum\jnl=14 Plasma Phys. Control. Fusion\ 
        \else\ifnum\jnl=15 Physiol. Meas.\ 
        \else\ifnum\jnl=16 Sov.\ Lightwave Commun.\
        \else\ifnum\jnl=17 J. Phys.\ D: Appl.\ Phys.\
        \else\ifnum\jnl=18 Supercond.\ Sci.\ Technol.\
        \else\ifnum\jnl=19 Semicond.\ Sci.\ Technol.\
        \else\ifnum\jnl=20 Nanotechnology\
        \else\ifnum\jnl=21 Meas.\ Sci.\ Technol.\ 
        \else\ifnum\jnl=22 Plasma Sources Sci.\ Technol.\ 
        \else\ifnum\jnl=23 Smart Mater.\ Struct.\ 
        \else\ifnum\jnl=24 J.\ Micromech.\ Microeng.\
   \else Institute of Physics Publishing\ 
   \fi\fi\fi\fi\fi\fi\fi\fi\fi\fi\fi\fi\fi\fi\fi
   \fi\fi\fi\fi\fi\fi\fi\fi\fi}
\let\abs=\beginabstract

\let\endabs=\endabstract
\def\today{\number\day\ \ifcase\month\or
     January\or February\or March\or April\or May\or June\or
     July\or August\or September\or October\or November\or
     December\fi\space \number\year}
\def\date{\ifppt\noindent\textfonts\rm 
     Date: \today\par\goodbreak\bigskip\fi}
%
%
\def\pacs#1{\ifppt\noindent\textfonts\rm 
     PACS number(s): #1\par\bigskip\fi}
%

%
%
\def\section#1{\ifppt\ifnum\secno=0\eject\fi\fi
    \subno=0\subsubno=0\global\advance\secno by 1
    \gdef\labeltype{\seclabel}\ifnumbysec\countno=1\fi
    \goodbreak\beforesecspace\nobreak
    \noindent{\bf \the\secno. #1}\par\futurelet\next\sp@ce}
\def\subsection#1{\subsubno=0\global\advance\subno by 1
     \gdef\labeltype{\seclabel}%
     \ifssf\else\goodbreak\beforesubspace\fi
     \global\ssffalse\nobreak
     \noindent{\it \the\secno.\the\subno. #1\par}%
     \futurelet\next\subsp@ce}
\def\subsubsection#1{\global\advance\subsubno by 1
     \gdef\labeltype{\seclabel}%
     \ifssf\else\goodbreak\beforesubsubspace\fi
     \global\ssffalse\nobreak
     \noindent{\it \the\secno.\the\subno.\the\subsubno. #1}\null. 
     \ignorespaces}
%

%
%
\def\numappendix#1{\ifappendix\ifnumbysec\countno=1\fi\else
    \countno=1\figno=0\tabno=0\fi
    \subno=0\global\advance\appno by 1
    \secno=\appno\gdef\applett{A}\gdef\labeltype{\seclabel}%
    \global\appendixtrue\global\numapptrue
    \goodbreak\beforesecspace\nobreak
    \noindent{\bf Appendix \the\appno. #1\par}%
    \futurelet\next\sp@ce}
\def\numsubappendix#1{\global\advance\subno by 1\subsubno=0
    \gdef\labeltype{\seclabel}%
    \ifssf\else\goodbreak\beforesubspace\fi
    \global\ssffalse\nobreak
    \noindent{\it A\the\appno.\the\subno. #1\par}%
    \futurelet\next\subsp@ce}
\def\@ppendix#1#2#3{\countno=1\subno=0\subsubno=0\secno=0\figno=0\tabno=0
    \gdef\applett{#1}\gdef\labeltype{\seclabel}\global\appendixtrue
    \goodbreak\beforesecspace\nobreak
    \noindent{\bf Appendix#2#3\par}\futurelet\next\sp@ce}
\def\Appendix#1{\@ppendix{A}{. }{#1}}
\def\appendix#1#2{\@ppendix{#1}{ #1. }{#2}}
\def\App#1{\@ppendix{A}{ }{#1}}
\def\app{\@ppendix{A}{}{}}
\def\subappendix#1#2{\global\advance\subno by 1\subsubno=0
    \gdef\labeltype{\seclabel}%
    \ifssf\else\goodbreak\beforesubspace\fi
    \global\ssffalse\nobreak
    \noindent{\it #1\the\subno. #2\par}%
    \nobreak\subspace\noindent\ignorespaces}
%
%
\def\@ck#1{\ifletter\bigskip\noindent\ignorespaces\else
    \goodbreak\beforesecspace\nobreak
    \noindent{\bf Acknowledgment#1\par}%
    \nobreak\secspace\noindent\ignorespaces\fi}
\def\ack{\@ck{s}}
\def\ackn{\@ck{}}
\def\n@ip#1{\goodbreak\beforesecspace\nobreak
    \noindent\smallfonts{\it #1}. \rm\ignorespaces}
\def\naip{\n@ip{Note added in proof}}
\def\na{\n@ip{Note added}}

%
%

%

%
%
\def\Tables{\vfill\eject\global\appendixfalse\textfonts\rm
    \everypar{}\noindent{\bf Tables and table captions}\par
    \bigskip}
\def\table#1{\tablecaption{#1}}
\def\tablecont{\topinsert\global\advance\tabno by -1
    \tablecaption{(continued)}}
\def\tablecaption#1{\gdef\labeltype{\tablabel}\global\widefalse
    \leftskip=\secindent\parindent=0pt
    \global\advance\tabno by 1
    \smallfonts{\bf Table \ifappendix\applett\fi\the\tabno.} \rm #1\par
    \smallskip\futurelet\next\t@b}
\def\endtable{\vfill\goodbreak}
\def\t@b{\ifx\next*\let\next=\widet@b
             \else\ifx\next[\let\next=\fullwidet@b
                      \else\let\next=\narrowt@b\fi\fi
             \next}
\def\widet@b#1{\global\widetrue\global\notfulltrue
    \t@bwidth=\hsize\advance\t@bwidth by -\secindent} 
\def\fullwidet@b[#1]{\global\widetrue\global\notfullfalse
    \leftskip=0pt\t@bwidth=\hsize}                  
\def\narrowt@b{\global\notfulltrue}
\def\align{\catcode`?=13\ifnotfull\moveright\secindent\fi
    \vbox\bgroup\halign\ifwide to \t@bwidth\fi
    \bgroup\strut\tabskip=1.2pc plus1pc minus.5pc}
\def\endalign{\egroup\egroup\catcode`?=12}

%
%
\def\L#1{#1\hfill}

\def\C#1{\hfill#1\hfill}
%
%
\def\br{\noalign{\vskip2pt\hrule height1pt\vskip2pt}}
\def\mr{\noalign{\vskip2pt\hrule\vskip2pt}}
%

%
%

%

\catcode`?=13
\def\lineup{\setbox0=\hbox{\smallfonts\rm 0}%
    \digitwidth=\wd0%
    \def?{\kern\digitwidth}%
    \def\\{\hbox{$\phantom{-}$}}%
    \def\-{\llap{$-$}}}
\catcode`?=12
%
%
\def\sidetable#1#2{\hbox{\ifppt\hsize=18pc\t@bwidth=18pc
                          \else\hsize=15pc\t@bwidth=15pc\fi
    \parindent=0pt\vtop{\null #1\par}%
    \ifppt\hskip1.2pc\else\hskip1pc\fi
    \vtop{\null #2\par}}} 
\def\lstable#1#2{\everypar{}\tempval=\hsize\hsize=\vsize
    \vsize=\tempval\hoffset=-3pc
    \global\tabno=#1\gdef\labeltype{\tablabel}%
    \noindent\smallfonts{\bf Table \ifappendix\applett\fi
    \the\tabno.} \rm #2\par
    \smallskip\futurelet\next\t@b}
\def\inctabno{\global\advance\tabno by 1}
%
%
\def\Figures{\vfill\eject\global\appendixfalse\textfonts\rm
    \everypar{}\noindent{\bf Figure captions}\par
    \bigskip}
\def\figure#1{\figc@ption{#1}\bigskip}
\def\figc@ption#1{\global\advance\figno by 1\gdef\labeltype{\figlabel}%
   {\parindent=\secindent\smallfonts\hang
    {\bf Figure \ifappendix\applett\fi\the\figno.} \rm #1\par}}
%
%
\def\refHEAD{\goodbreak\beforesecspace
     \noindent\textfonts{\bf References}\par
     \let\ref=\rf
     \nobreak\smallfonts\rm}
\def\numreferences{\refHEAD\parindent=30pt
     \everypar{\hang\noindent\frenchspacing\rm}
     \secspace}
\def\rf#1{\par\noindent\hbox to 21pt{\hss #1\quad}\ignorespaces}
%

%

%
%

%
%

%
%

%
%

%
\catcode`\@=12
%
%
\def\pptstyle{\ppttrue\headsize{17}{24}%
\textsize{12}{16}%
\smallsize{10}{12}%
\hsize=37.2pc\vsize=56pc
\textind=20pt\secindent=6pc}
%
%

%
%

%
%

%
\parindent=\textind
\input xref
\def\ie{i.e.,}
\def\eg{e.g.,}
\def\ea{{\it et al}}

\pptstyle
\jl{3}
\title
{Density-Functional Theory of Quantum Freezing:
Sensitivity to Liquid-State Structure and Statistics}[Density-Functional 
Theory of Quantum Freezing]
\author
{A R Denton\dag\footnote{*}{Present address: 
Institut f\"ur Festk\"orperforschung, Forschungszentrum J\"ulich GmbH, 
D-52425 J\"ulich, Germany (a.denton@kfa-juelich.de)},~~P Nielaba\ddag~~and
N W Ashcroft\S}[A R Denton et al]
\address
{\dag\ Institut f\"ur Theoretische Physik, Technische Universit\"at Wien, 
Wiedner Hauptstra{\ss}e 8-10, A-1040 Wien, Austria}
\address
{\ddag\ Institut f\"ur Physik, Johannes Gutenberg Universit\"at,\hfil\break 
Staudinger Weg 7, D-55099 Mainz, Germany}
\address
{\S\ Laboratory of Atomic and Solid State Physics and Materials Science Center, 
Cornell University, Ithaca, NY 14853-2501, USA}
\abs
Density-functional theory is applied to compute the ground-state energies
of quantum hard-sphere solids.  The modified weighted-density approximation
is used to map both the Bose and the Fermi solid onto a corresponding 
uniform Bose liquid, assuming negligible exchange for the Fermi solid.
The required liquid-state input data are obtained from a paired phonon 
analysis, combined with an enhanced hypernetted-chain integral equation, 
and the Feynman approximation, connecting the static structure factor 
and the linear response function. 
The Fermi liquid is treated by the Wu-Feenberg cluster expansion, which
approximately accounts for the effects of antisymmetry.
Liquid-solid transitions for both systems are obtained with no adjustment of
input data, the Fermi liquid freezing at lower density than the Bose liquid
because of the destabilizing influence of fermionic statistics.
Predictions for these quantum systems are shown to be more sensitive 
to the accuracy of the input data than is the case for the classical 
counterpart.  Limited quantitative agreement with simulation indicates 
a need for still further improvement of the liquid-state input, likely 
through practical alternatives to the Feynman approximation.  
\endabs

\date
\pacs{64.70.Dv, 05.70.-a, 64.60.-i, 67.80.-s}

\section{Introduction}
Modern density-functional (DF) theory of non-uniform systems~\cite{DF1}
continues to yield useful insight into the fundamental nature 
of the liquid-solid (freezing) transition, 
as well as practical approximations for 
basic thermodynamic properties of the solid phase.
Treating the solid as a highly non-uniform system, and taking as input 
known structural and thermodynamic properties of the corresponding uniform 
system, the DF approach provides an approximation for the free energy of 
a solid of prescribed symmetry, as a functional of the one-particle density.
From the solid free energy follow freezing parameters, such as the densities 
of coexisting liquid and solid phases and the Lindemann ratio, as well as 
various solid-state properties, including the equation of state and relative 
stabilities of competing structures.

The Ramakrishnan-Yussouff (RY)~\cite{RY} perturbative DF theory of freezing, 
which is based on a truncated functional Taylor-series expansion of the 
non-uniform system grand potential about that of the uniform liquid, 
initiated a number of subsequent reformulations.  Among these are several 
non-perturbative theories, which posit a thermodynamic or structural 
``mapping" of the non-uniform system onto the uniform liquid.  
Although differing to varying extents in both philosophy and manner of 
approximation, all formulations of the theory require the same essential 
input, namely the direct correlation function (DCF) of the uniform liquid.  
For classical systems, the latter is trivially related
to the static structure factor, which is directly obtainable from 
integral equation theories, simulations, or scattering experiments~\cite{HM}.
Numerous applications of DF theory to a wide range of bulk and interfacial 
systems have clearly demonstrated its utility as a practical tool for 
studying classical non-uniform system phenomena~\cite{DF2}.

In recent years, extensions of the DF approach to quantum systems at
zero temperature have also been developed and applied to freezing of various 
ground-state systems~\cite{QDF}, including $^4$He~\cite{MS1,Dalfovo}, 
electrons~\cite{SP,MS2,CG} 
(Wigner crystallization of the Fermi one-component plasma), and
Bose hard spheres~\cite{DNA}.
For finite temperatures, a path-integral-based extension of the (classical)
RY theory~\cite{MRH,RMH} has been proposed and applied
to freezing of helium.  However, a general theory capable of describing 
freezing both in the ground state {\it and} at finite temperatures is 
still not at hand.

Ground-state quantum DF theory is formally similar to the classical theory, 
but a major difference is that it requires as input the quantum analog
of the DCF of the uniform system.  
A significant complication for the quantum theory is the lack of 
any simple relationship between the {\it quantum} DCF
[defined below by (8) and (9)] and the static structure factor.
Since the latter is generally much easier to compute, in practice it is 
usually necessary to resort to approximate relationships.
Unfortunately, this has tended to blur the distinction between the issues 
of the accuracy of the theory and the accuracy of the input to the theory.
Recently, powerful quantum Monte Carlo (MC) methods have been employed to 
directly compute quantum DCFs for certain 
systems~\cite{CS1,CS2}, thereby raising the hope that more definitive 
tests of DF theory will soon be possible.  Nevertheless, it remains 
worthwhile to investigate the utility of quantum DF theory using as input 
the currently more accessible, if somewhat less accurate, liquid-state 
structural data.

In previous work~\cite{DNA}, we extended one version of DF theory, based 
on the modified weighted-density approximation (MWDA)~\cite{MWDA}, from 
classical to ground-state quantum systems and demonstrated its application
to freezing of the Bose hard-sphere (HS) liquid at zero temperature.  
For the liquid-state input data, we used an iterative procedure based on 
({\it i}) the paired phonon analysis (PPA)~\cite{PPA} to 
solve the Euler-Lagrange equation for the optimum pair pseudopotential, 
({\it ii}) the hypernetted-chain (HNC) integral equation~\cite{HM} to 
compute the corresponding ground-state energy and static structure factor, and
({\it iii}) the Feynman approximation to connect the static structure factor 
to the quantum DCF.
In an attempt to compensate for inaccuracies in the input data, we simply 
scaled the DCF by a constant factor at all wave vectors.  
Despite a favourable comparison of predicted solid ground-state energies and 
freezing parameters with available simulation data, the sensitivity of the 
results to the scaling {\it ansatz} remained unclear.

The main purpose of this paper is to make use of significantly more accurate 
data for the energy and structure of the Bose HS liquid to specifically 
examine the sensitivity of quantum DF theory to the quality of the 
liquid-state input data.
In the process, we also examine the qualitative influence of 
particle statistics by considering freezing of both Bose and Fermi HS liquids.
Aside from being of fundamental interest as idealized model systems
in which effects of excluded volume interactions may be studied in isolation, 
HS systems also serve as useful reference systems for perturbation-theory 
descriptions of real quantum systems, such as helium and dense neutron matter.
In Sec.~2 we begin by
reviewing the principles of DF theory and the MWDA in the context of 
ground-state quantum systems.  Section 3 concerns the structure of the 
liquid and describes ({\it i}) calculation of the static structure factor 
for Bose hard spheres via the PPA and an enhanced HNC integral equation
that includes four- and five-body elementary diagrams~\cite{KS}, ({\it ii})
approximate connections between the static structure factor and the quantum 
DCF, and ({\it iii}) adjustment of the Bose liquid energy 
for Fermi statistics using the cluster expansion method of 
Wu and Feenberg~\cite{WF}.
In contrast to our previous study, {\it no scaling or any other modification 
of input data is now employed}.  In Sec.~4 we outline our application of DF 
theory to freezing of quantum liquids and present results for hard spheres.  
Considerable improvement in the ground-state solid energies results from 
use of the more accurate liquid-state data, and distinct freezing transitions 
are obtained for both Bose and Fermi systems. 
Quantum statistics are found to significantly influence the 
freezing parameters, the energetically less stable Fermi liquid 
crystallizing at lower density, and with weaker atomic localization, 
than the Bose liquid. 
Quantitative discrepancies between theory and simulation
are discussed in the context of the sensitivity 
of the theory to the accuracy of the input data. 
In Sec.~5 we summarize and conclude by suggesting the need for 
further improvement on the side of liquid-state theory.  
Finally, in the Appendix we examine the exact short-wavelength asymptotic 
behaviour of the quantum DCF and possibly important implications 
for the accuracy of the Feynman approximation.

\section{Density-Functional Theory}
At zero temperature the central quantity of interest in DF theory is the total 
ground-state energy $E[\rho]$, a unique functional of the (spatially-varying) 
density $\rho({\bdi r})$ that is minimized by the equilibrium 
density~\cite{HK}, {\ie}
$${{\delta E[\rho]}\over{\delta\rho({\bdi r})}} = 0.\label{var}$$
In general, $E[\rho]$ may be separated, according to
$$E[\rho]=E_{id}[\rho]+E_c[\rho]+E_{ext}[\rho],\label{E}$$
into an ideal-gas energy $E_{id}[\rho]$ (energy of the non-uniform 
system in the absence of interactions), a correlation energy $E_c[\rho]$ 
due to internal interactions and exchange, and an external energy 
$$E_{ext}[\rho] = \int d{\bdi r} \rho({\bdi r}) \phi_{ext}({\bdi r})
\label{Eext}$$
due to interaction with an external potential $\phi_{ext}({\bdi r})$.
One advantage of this separation is that for bosons $E_{id}[\rho]$ is given by 
the exact expression~\cite{DNA}
$$E_{id}[\rho]={\hbar^2\over{2m}}{\int}d{\bdi r}|
\bnabla\sqrt{\rho({\bdi r})}|^2,\label{Eid1}$$
where $m$ is the particle mass. 
Another advantage is that $E_c[\rho]$, although not known exactly for 
non-uniform systems, is amenable to approximation by any of the standard DF
approximations applied to the classical excess free energy.  Here, as in
previous work~\cite{DNA}, we approximate $E_c[\rho]$ by an extension to 
ground-state quantum systems of the modified weighted-density 
approximation~\cite{MWDA}.

The MWDA is based on the general assumption that the non-uniform system 
may be mapped onto a suitably chosen uniform system~\cite{WDA}.  In the 
context of ground-state quantum systems, this implies that the 
average correlation energy per particle of the non-uniform system is 
equated to that of a uniform liquid $\epsilon$, but evaluated at an 
effective density $\hat\rho$, {\ie}
$$E_c^{MWDA}[\rho]/N{\equiv}\epsilon(\hat\rho),\label{Ec}$$
where $N$ is the number of particles. 
For classical solids, as well as quantum solids obeying Bose statistics, 
the natural choice of uniform system for the mapping is the corresponding
liquid.  In the case of Fermi systems, however, the effect of particle 
statistics is known to be quite different in the liquid and solid phases.
In particular, the exchange contribution to the energy is considerably 
larger in the uniform liquid than it is in the solid, in which exchanges 
amongst site-localized atoms are relatively rare.  
Comparisons of MC results for solid $^3$He and $^4$He~\cite{HLS,CCK}, 
for example, show a negligible effect of statistics on ground-state energies.
Furthermore, the energies for solid $^3$He as calculated by 
Ceperley {\ea}~\cite{CCK}, using a MC method to sample the square of
a properly antisymmetrized wave function, are in good agreement with
the VMC results of Hansen and Levesque~\cite{HLS}, who used an
unsymmetrized Jastrow trial wavefunction (see Sec.~3).
These considerations motivate our assumption that within the MWDA 
the Fermi solid, like its Bose counterpart, is best mapped onto 
the corresponding Bose liquid.
This is equivalent to ignoring the symmetry with respect to particle exchange
of the solid many-body wavefunction, on the grounds that atoms in the solid 
are sufficiently localized about identifiable lattice sites as to be 
essentially distinguishable. 
The physical interpretation of the MWDA for quantum solids is thus 
the following:
The correlation part of the ground-state energy of both the Bose {\it and} 
Fermi solids is represented by that of the corresponding Bose liquid 
of appropriate effective density.
As this effective density is significantly lower in practice than 
the average solid density (see Sec.~4), the solid correlation energy 
is in fact modelled by that of a relatively low-density liquid, 
thus properly reflecting the fact that interatomic correlations are 
reduced upon confinement of the atoms to equilibrium lattice sites.
Clearly in the case of more weakly inhomogeneous quantum systems, such as 
thin films or liquids near interfaces, neglect of exchange effects would be 
inadequate.  An interesting open issue for future consideration is then 
how quantum DF theory might be generalized to describe Fermi systems of 
{\it arbitrary} inhomogeneity.

The effective or {\it weighted} density $\hat\rho$ is now
assumed to depend on a weighted average over the volume of 
the system of the physical density, and is defined by
$$\hat\rho{\equiv}{1\over{N}}{\int}d{\bdi r}\rho({\bdi r}){\int}d{\bdi r}'
\rho({\bdi r}')w({\bdi r}-{\bdi r}';\hat\rho),\label{wd1}$$
with $\hat\rho$ appearing also implicitly as the argument of the 
weight function $w$.  Normalization of $w$, according to
$${\int}d{\bdi r}'w({\bdi r}-{\bdi r}';\rho) = 1,\label{norm}$$
ensures that the approximation is exact in the limit of a uniform liquid 
[$\rho({\bdi r}) \to \rho$].  Unique specification of $w$ follows from 
further requiring that $E_c^{MWDA}[\rho]$ generate the exact 
``particle-hole interaction"~\cite{KS} in the uniform limit, {\ie} 
$$\lim_{\rho({\bdi r}) \to \rho}\Bigl({{{\delta
^2E_c^{MWDA}[\rho]}\over{\delta\rho({\bdi r})\delta\rho({\bdi r}')}}}\Bigr)
= v(|{\bdi r}-{\bdi r}'|;\rho),\label{vr}$$
where $v(|{\bdi r}-{\bdi r}'|;\rho)$ is to be interpreted as the 
quantum analog of the classical Ornstein-Zernike two-particle DCF, 
henceforth referred to as the {\it quantum} DCF.
Formally, equation \ref{vr} ensures that a functional Taylor-series expansion 
of $E_c^{MWDA}[\rho]$ about
the density of a uniform reference liquid is exact to second order, and
also includes approximate terms to all higher orders~\cite{MWDA,WDA}.

A relationship that proves useful in approximating the quantum DCF
(see Sec.~3) is obtained by taking two functional derivatives of 
\ref{E} and using \ref{var}, \ref{Eext}, and \ref{Eid1}, yielding
$$\fl v(|{\bdi r}-{\bdi r}'|;\rho) = -{{\delta\phi_{ext}({\bdi r})}\over
{\delta\rho({\bdi r}')}} - {{\delta^2 E_{id}[\rho]}\over{\delta\rho({\bdi r})
\delta\rho({\bdi r'})}} = -\Bigl[\chi^{-1}(|{\bdi r}-{\bdi r}'|;\rho)-
\chi_o^{-1}(|{\bdi r}-{\bdi r}'|;\rho)\Bigr],\label{vchi}$$
where
$$\chi^{-1}(|{\bdi r}-{\bdi r}'|;\rho) = 
{{\delta\phi_{ext}({\bdi r})}\over{\delta\rho({\bdi r'})}}\label{chi}$$
is the functional inverse of the static density-density 
linear response function and
$$\chi_o^{-1}(|{\bdi r}-{\bdi r}'|;\rho) = -{{\delta^2 E_{id}[\rho]}\over
{\delta\rho({\bdi r})\delta\rho({\bdi r'})}}.\label{chio1}$$
is its non-interacting (free-particle) limit.  The Fourier transform 
of the latter can be expressed [using \ref{Eid1}] in the simple form
$$\chi_o^{-1}(k) = -{{\hbar^2 k^2}\over{4m\rho}}.\label{chio2}$$

In Fourier space, where computations are most conveniently carried out, 
the weight function is given by the analyic relation
$$w(k;\rho)={1\over{2\epsilon'(\rho)}}\Bigl[v(k;\rho)
-\delta_{k,0}\rho\epsilon''(\rho)\Bigr],\label{wk}$$
which follows directly from \ref{Ec}-\ref{vr}.
Note that normalization of the weight function implies, 
via \ref{wk}, the quantum compressibility sum rule 
$$v(k=0;\rho)=2\epsilon'(\rho)+\rho\epsilon''(\rho)=mc^2/\rho,\label{cr}$$
$c$ being the speed of longitudinal sound.  This sets a consistency 
constraint on the liquid-state energy and structure.

In summary, equations \ref{Ec}, \ref{wd1}, \ref{norm}, and \ref{wk} constitute 
the MWDA for the correlation energy of a quantum solid at zero temperature.
Practical implementation of the theory requires knowledge of the liquid-state 
correlation energy $\epsilon$ and the quantum DCF $v$, 
the subject of the next section.

\section{Liquid-State Energy and Structure}
For the ground-state Bose HS system, the liquid-state energy and 
static structure factor have been obtained by a method that combines the 
paired phonon analysis (PPA)~\cite{PPA} with integral equation theory.
The extraction of the key quantum DCF from the 
static structure factor is described in detail below.  
First, however, we briefly summarize the PPA.

The basis of the PPA, as well as of variational Monte Carlo (VMC) methods, 
is a trial ground-state wavefunction of the general Feenberg form
$$ \psi_B({\bdi r}_1,\ldots,{\bdi r}_N)
=\exp[-\sum_{i<j}u_2({\bdi r}_i,{\bdi r}_j)
-\sum_{i<j<k}u_3({\bdi r}_i,{\bdi r}_j,{\bdi r}_k)-\cdots], \label{trial1}$$
from which the optimum correlation factors $u_n$ are determined by 
minimization of the total ground-state energy, according to
$$ { \delta \over { \delta u_n } }
      \left[
      { { \langle \psi_B | H | \psi_B \rangle }
        \over
        { \langle \psi_B |  \psi_B \rangle } }
      \right] = 0,
       \label{ufunct} $$
where $H$ is the Hamiltonian.
In order to be able to directly compare with results of VMC simulations, 
we mainly concern ourselves here with a trial wavefunction of the Jastrow form, 
which includes only the pair correlation factor, or pseudopotential,  
$u_2(r)\equiv u(r)$.  We have also considered, however, the effect of 
including triplet correlations and briefly comment on this in Sec.~4.
The pseudopotential is constrained by the general condition that 
$u(r){\to}0$ as $r\to\infty$, implying vanishing correlations at infinite
separation, and the condition (specific to hard spheres of diameter $\sigma$) 
that $u(r)\to\infty$ as $r\to\sigma$, ensuring a minimum pair
separation of $\sigma$.
In applications to solids, VMC methods assume \ref{trial1} multiplied by 
a product of one-body (typically Gaussian) wavefunctions, 
each centred on a lattice site.

By exploiting the formal correspondence between the quantum probability 
density $|\psi_B|^2$ and the classical Boltzmann factor~\cite{McM}, in which 
$u(r)$ may be interpreted as a classical pair potential, 
the method uses integral equation theory for {\it classical} liquids 
to determine the radial distribution function $g(r,[u])$ for a given $u(r)$.
In previous work~\cite{DNA}, we used the approximate HNC integral equation, 
which entirely neglects the bridge function.  Here, we use a more accurate 
enhancement of the HNC equation (denoted by ``HNC+") that 
approximates the bridge function by four- and five-body elementary diagrams.
From $g(r,[u])$, the corresponding correlation energy is computed
via~\cite{PPA,McM}
$$E_c(\rho) = {1\over{2}}\rho\int d{\bdi r}g(r,[u])
[{\hbar^2\over{2m}}\nabla^2u(r)+\phi(r)],\label{gu}$$
where $\phi(r)$ is the pair potential (for hard spheres in our case).
The random phase approximation is then used iteratively to adjust the 
pseudopotential until the energy is minimized.
Figure~1 illustrates the resulting static structure factor $S(k)$, which
is related to $g(r)$ by the Fourier transform relation
$$S(k) = 1+\rho\int d{\bdi r} [g(r)-1] \exp(i{\bdi k}\cdot{\bdi r}).
\label{FT}$$

To date there is no known exact relation connecting $S(k)$ to the 
quantum DCF $v(k)$.  However, as shown in Sec.~2, 
$v(k)$ is trivially related to the static response function $\chi(k)$.  
Further, the fact that $S(k)$ and $\chi(k)$ may be expressed as 
frequency moments
$$m_p(k) = \int_0^\infty d\omega (\hbar\omega)^p S(k,\omega),\label{mp}$$
of the dynamic structure factor $S(k,\omega)$ 
provides a basis for approximations~\cite{HF}.
The relevant lowest-order moments (or sum rules) are given explicitly by
$$m_{-1} = -{1\over{2\rho}}\chi(k),\label{m-a}$$
$$m_0 = S(k),\label{m-b}$$
and
$$m_1 = {{\hbar^2k^2}\over{2m}}.\label{m-c}$$
From the rigorous inequality
$$\int_0^{\infty}d\omega {{S(k,\omega)}\over{\hbar\omega}}
(1+a\hbar\omega)^2 = m_{-1}+2a m_0+a^2m_1 \geq 0,\label{ineq1}$$
variation of the real parameter $a$ to minimize the 
left-hand side yields the bound~\cite{HF}
$$m_{-1}(k) \geq {{2mS^2(k)}\over{\hbar^2k^2}}.\label{mbound1}$$
The treatment of \ref{mbound1} as an equality, rather than merely as an 
upper bound, is the so-called Feynman approximation,
$$m^F_{-1}(k) = {{2mS^2(k)}\over{\hbar^2k^2}}.\label{Feynmanm}$$
From \ref{vchi}, \ref{chio2}, and \ref{m-a}, the corresponding 
Feynman approximation for $v(k)$ is
$$v^F(k) = {{\hbar^2k^2}\over{4m\rho}}\Bigl({1\over{S^2(k)}}-1\Bigr),
\label{Feynmanv}$$
which is plotted in Figure~2 for Bose hard spheres, using the PPA $S(k)$.

In the long-wavelength ($k \to 0$) limit, the Feynman approximation 
evidently correctly obeys \ref{cr}, since $S(k)$ is known to vanish 
linearly with $k$, according to~\cite{CR}
$$S(k) \sim {{\hbar}k\over{2mc}} \quad (k \to 0).\label{smallk}$$
However, the PPA -- in common with all liquid theories -- does not satisfy
the compressibility sum rule [equation \ref{cr}] relating $\epsilon$ 
and $v(k=0)$.
That is, the speed of longitudinal sound derived from $\epsilon$
differs from that derived from $v(k=0)$.
At $k=0$ the discrepancy is unimportant, since the MWDA satisfies the 
sum rule by construction [see \ref{norm} and \ref{wk}].  
At $k\neq 0$ any remnant of the discrepancy is important only to the extent 
that $v(k)$ may be affected at relevant wave vectors, {\eg} the reciprocal 
lattice vectors of the crystal in the application to freezing (Sec.~4). 
Recently a simple analytic modification of $v(k)$ 
has been introduced~\cite{CFK} to correct the low-$k$ behaviour.
This correction, however, would have negligible effect on $v(k)$ at the 
wave vectors relevant to freezing, since the first reciprocal lattice vector 
of the crystal generally lies at or beyond the first minimum in $v(k)$ 
(see Figure~2).
Thus, although in previous work~\cite{DNA} the sum rule violation motivated 
an empirical scaling of $v(k)$, in the present work we make 
{\it no alteration to the PPA input}.

In the opposite short-wavelength ($k \to \infty$) asymptotic limit, 
it may be shown from \ref{Feynmanv} that $v^F(k) \to 0$, 
which is identical to the corresponding limit of the classical DCF.
In the Appendix we consider a recently-proposed extension to the 
Feynman approximation, involving higher-order moments of $S(k,\omega)$, 
and use this to analyse the {\it exact} asymptotic behaviour of $v(k)$.  
Interestingly, unlike its classical counterpart, the exact quantum DCF 
tends not to zero, but rather to a density-dependent constant.
The same property has been discussed recently, in another context, 
by Likos et al~\cite{CL}, and may have important implications for 
the form of the exact quantum DCF in the finite-$k$ range 
of relevance for the DF theory of quantum solids.

For the Fermi HS liquid, we have approximated the ground-state energy 
by adjusting the Bose liquid energy for Fermi statistics using the 
Wu-Feenberg (or Pandharipande-Bethe) cluster expansion method~\cite{WF}.
The same approach has been adopted previously in VMC studies of liquid 
$^3$He~\cite{HLS} and of the Fermi HS liquid~\cite{Schiff}.
The Wu-Feenberg (WF) method assumes that the
Pauli exclusion principle acts only as a weak perturbative correction to the
strongly repulsive pair potential, and it takes as a trial wavefunction 
$\psi_F$ the Bose wavefunction $\psi_B$ antisymmetrized by a 
Slater determinant:
$$ \psi_F({\bdi r}_1,\ldots,{\bdi r}_N)
     =\psi_B({\bdi r}_1,\ldots,{\bdi r}_N)\cdot\det
[\exp(i{\bdi k}_i\cdot{\bdi r}_j)].\label{trial2}$$
The corresponding ground-state energy per particle $\epsilon^F$ may be 
expressed as a cluster expansion about the Bose counterpart $\epsilon^B$, 
according to
$$\epsilon^F=\epsilon^B+\epsilon_0^F+\epsilon_1^F+\epsilon_2^F,
\label{WF-a}$$
where
$$\epsilon_0^F={3\over{5}}\epsilon_F,\label{WF-b}$$
$$\epsilon_1^F=24\epsilon_F\int_0^1 dx x^4(1-{3\over{2}}x+{1\over{2}}x^3)
[S(2k_Fx)-1],\label{WF-c}$$
and
$$\fl \epsilon_2^F=-\Bigl({3\over{8\pi}}\Bigr)^3\epsilon_F\int d{\bdi x}_1
\int d{\bdi x}_2 \int d{\bdi x}_3 x_{12}^2 S(k_Fx_{12})
[S(k_Fx_{23})-1] [S(k_Fx_{31})-1],\label{WF-d}$$
with $k_F=(3\pi^2\rho)^{1/3}$ and $\epsilon_F=(\hbar^2k_F^2/2m)$ being the 
Fermi wave-vector and energy, respectively, and with the integrals in 
\ref{WF-d} extending over the unit sphere.
Evidently $\epsilon_0^F$ is the one-particle correction arising directly 
from the Pauli exclusion principle for non-interacting fermions, while 
$\epsilon_1^F$ and $\epsilon_2^F$ arise from antisymmetrization of 
two- and three-particle clusters, respectively.
We have computed the first-order correction $\epsilon_1^F$ 
by standard numerical quadrature and the second-order correction
$\epsilon_2^F$ by a simple Monte Carlo integration algorithm that randomly 
chooses triplets of points within the nine-dimensional integration volume.
Figure~3 shows the resulting relative magnitudes of successive terms
in the cluster expansion.  

At first glance, the apparent rapid convergence of the first few terms
in the WF expansion would seem to lend some {\it a posteriori} justification 
to the assumption of a small perturbation.
Serious doubts have been raised, however, concerning the general
convergence properties of such cluster expansions~\cite{CCK,Brandow,krot1}.
Brandow~\cite{Brandow} has identified two distinct antisymmetry effects
and argued that one of these, the so-called ``correlation-exclusion" effect, 
is treated in a non-converging manner, and in fact can be properly accounted 
for only by including diagrams involving exchanges amongst any number of 
particles up to and including all $N$ particles.
The MC calculations of Ceperley {\ea}, for a variety of pair-wise-interacting
Fermi systems, demonstrate that the WF expansion tends to 
consistently underestimate the energy, particularly for soft-core 
pair potentials and at high densities. 
Krotscheck~\cite{krot2} has shown that conventional cluster expansion
methods tend to be unreliable in the case of high densities and long-ranged
Jastrow functions, and stressed the importance of summing to all orders 
certain properly combined classes of diagrams (Fermi chains).
The Fermi HNC method, although summing much larger classes of diagrams than
the WF expansion, also is plagued by uncertain convergency~\cite{krot2}.
Despite the somewhat doubtful convergence properties of the WF expansion, 
we nevertheless employ it here, pending a more accurate treatment of the
Fermi liquid~\cite{KD}, in order to make contact with the available VMC 
simulation data~\cite{Schiff}, and to estimate -- at least qualitatively -- 
the influence of Fermi statistics on the freezing parameters of hard spheres.

\section{Freezing of Quantum Hard Spheres}
The main steps in applying DF theory to bulk solids and freezing transitions 
are parametrization of the solid density, minimization of the total solid 
energy with respect to the parametrized density, and identification of 
liquid-solid coexistence by construction of a Maxwell common tangent.
Below we outline the procedure, which is described in greater detail 
elsewhere~\cite{DNA}.

Assuming a Gaussian density distribution about the lattice sites of a perfect 
{\it fcc} crystal, the solid density takes the form
$$\rho_s({\bdi r})=({\alpha\over{\pi}})^{3\over 2}\sum_{\bdi R}
\exp(-\alpha|{\bdi r}-{\bdi R}|^2),\label{gauss}$$
where the localization parameter $\alpha$ determines the width of the 
Gaussians centred on the lattice sites at positions ${\bdi R}$.
This choice corresponds to including a Gaussian one-body factor in 
the trial wavefunction of \ref{trial1}.
We focus here on the {\it fcc} crystal structure, since solid helium is known 
from experiment to assume a close-packed structure at zero temperature and
corresponding pressures~\cite{Wilks}, and since our previous study indicated 
negligible differences between the relative stabilities of {\it fcc}, 
{\it hcp}, and {\it bcc} structures.
The Gaussian {\it ansatz} has been widely adopted in applications of 
DF theory to {\it classical} systems~\cite{DF1,DF2}.  Simulation
studies by Young and Alder~\cite{YA} and by Ohnesorge {\ea}~\cite{OLW} 
have demonstrated the global form of the classical HS crystal density
distribution to be close to Gaussian, especially near close packing, although
near melting the tails of the distribution do exhibit significant
anisotropic deviations of roughly $10$\% from the Gaussian form~\cite{OLW}.
The only comparable study for quantum systems that we are aware of is the
Green's Function Monte Carlo (GFMC) simulation of Whitlock {\ea}~\cite{WCCK}
for a Lennard-Jones model of $^4$He, which found a spherically symmetric 
density distribution about a given lattice site with only small positive 
deviations from Gaussian behaviour in the tail of the distribution.
Since freezing properties are known to be dominated by short-range
repulsive forces, which are similar in HS and Lennard-Jones 
systems, we can expect the Gaussian {\it ansatz} also to be a reasonable 
approximation for quantum hard spheres.  
Caution is warranted, however, in this application
to a relatively low-density quantum solid to the extent that the solid energy
may be sensitive to anisotropy and other deviations of the density from 
the simple Gaussian form.
A definitive resolution of the issue could be obtained by the technique
of free minimization~\cite{OLW}, which avoids entirely the need for density 
parametrization, although we have not attempted such an extensive computation.

From \ref{wk} and \ref{gauss}, \ref{wd1} now takes the form 
$$\hat\rho(\alpha,\rho_s)=\rho_s\Bigl[1+{1\over{2\epsilon'
(\hat\rho)}}\sum_{{\bdi G}\neq 0}\exp(-G^2/2\alpha)
v(G;\hat\rho)\Bigr],\label{wd3}$$
where $\rho_s$ is the average solid density and
$G$ the magnitude of the {\it fcc} reciprocal lattice vector ${\bdi G}$.
Given the liquid-state functions $\epsilon(\rho)$ and $v(k;\rho)$ from
Sec.~3, this implicit relation for $\hat\rho$ 
can be easily solved by numerical iteration at fixed $\alpha$ and $\rho_s$.
From $\hat\rho$ the approximate correlation energy $E_c^{MWDA}$ then
follows immediately from \ref{Ec}.

It is important to emphasize that here we map both the Bose and the
Fermi HS solids onto the corresponding Bose liquid, under the assumption that 
exchange effects in the Fermi solid are negligible.  As noted in Sec.~2, 
this assumption is expected to be valid in the solid,  
where overlap of neighbouring density distributions is negligible.  
In this case, the ideal-gas energy $E_{id}$ 
takes the simple analytic form~\cite{DNA} 
$$E_{id}/N \simeq {3\over{4}}{\hbar^2\over{m}}\alpha,\label{Eid2}$$
identical to the form usually assumed in VMC simulations~\cite{HLS}.
Assuming no external potential, \ref{Ec} and \ref{Eid2} combine finally 
to give the approximate total energy per particle
$$E^{MWDA}(\alpha,\rho_s)/N \simeq {3\over{4}}{\hbar^2\over{m}}\alpha+
\epsilon(\hat\rho(\alpha,\rho_s)).\label{EMWDA}$$
Note that in the case of helium the mass difference between the $^3$He 
and $^4$He isotopes gives rise to quite different kinetic contributions 
to the total energy.  For the purely kinetic HS systems considered here, 
however, the energies of the Bose and Fermi solids 
(in units of $\hbar^2/2m\sigma^2$) are identical by assumption.

Minimization of \ref{EMWDA} with respect to the single variational parameter
$\alpha$ (at fixed $\rho_s$) is illustrated in Figure~4, which shows separately 
the variation of the ideal-gas and correlation energies with $\alpha$.
The linear increase of $E_{id}$, which results from the kinetic energy 
cost of increasing the curvature of the one-particle wavefunction, strongly 
opposes localization of the atoms about their lattice sites.  In contrast,  
the rapid decrease of $E_c$, which arises from a weakening 
of interactions with reduced overlap of neighbouring density distributions,  
strongly favours localization.
At sufficiently high density, the competition between $E_{id}$ and $E_c$ 
results in a minimum in the {\it total} energy at non-zero $\alpha$, 
signalling mechanical stabilization of the solid.
By varying $\rho_s$ and repeating the above minimization procedure, the 
solid equation of state ($E$ vs $\rho_s$) is obtained. 
Figure~5 displays the corresponding weighted density, confirming the 
characteristic trait of the MWDA, referred to in Sec.~2, that 
$\hat\rho$ is always lower than the average solid density.
It should be noted that Figure~5 differs from the corresponding figure
in~\cite{DNA} (second reference) due to a plotting error in the latter, 
and that the roughness of the curve arises from linear interpolation 
between grid points of the liquid-state $v(k)$. 

Thermodynamic stability of the solid is assessed by comparing the 
liquid and solid energies.  
Coexistence between liquid and solid, characterized by equality of 
the pressures and of the chemical potentials in the two phases, 
may be established by constructing a common tangent -- {\it if one exists} 
-- to the energy vs. density curves,  the coexistence densities 
occurring at the points of common tangency.
The corresponding value of $\alpha$ determines the Lindemann ratio $L$, 
defined as the ratio of the {\it rms} atomic displacement to the 
nearest-neighbour distance in the solid at coexistence.  
For the {\it fcc} crystal $L=(3/\alpha a^2)^{1/2}$, where $a$ is the 
lattice constant.

Applying the above procedure, we have computed the energy as a function of
density for the ground-state quantum HS {\it fcc} crystal 
and have attempted a freezing analysis for the Bose and Fermi systems.  
In order to assess the sensitivity of the theory to the accuracy of the 
liquid-state input data, we have performed calculations for the Bose system 
using data generated by the PPA with both the simple HNC and the 
enhanced HNC+ approximations restricted to pair correlations (see Sec.~3). 
The PPA static structure factor -- and hence the quantum DCF
[from \ref{Feynmanv}] -- turns out to be practically identical 
regardless of whether the HNC or HNC+ closure is used. 
The energy, however, differs significantly.  
Figure~6 shows the Bose HS liquid energy density vs
number density in the two approximations, together with the corresponding 
{\it fcc} crystal energy density predicted by our DF theory.  
Also shown, for comparison, are the VMC simulation data of 
Hansen {\ea}~\cite{HLS}.
In both cases, the existence of a common tangent to the liquid and solid 
curves confirms the occurrence of a freezing transition.  
The corresponding
freezing parameters are given in Table~1, where it is seen that 
in comparison with simulation the theory considerably overestimates 
the coexistence densities as well as the density change, and 
underestimates the Lindemann ratio. 
Evidently, use of the more accurate HNC+ approximation significantly 
improves agreement with simulation for both the liquid and solid energies.  
In fact, the liquid energy agrees almost perfectly up to $\rho\sigma^3=0.4$, 
well beyond the freezing density.
Nevertheless, the coexistence densities are slightly more accurate 
within the HNC approximation, highlighting the delicate dependence 
of the freezing parameters on the detailed form of the equation of state.  
Despite quantitative discrepancies in the predicted freezing parameters, 
it is important to point out that the prediction of a freezing transition 
itself is a significant qualitative result, considering that the 
second-order RY version of DF theory gives no freezing transition 
for the same input data~\cite{MS1,Dalfovo}.

To test for sensitivity of the results to the assumption of pair correlations 
in the trial wavefunction of \ref{trial1}, we have repeated the calculations 
using PPA input data that include triplet correlations. 
Although the liquid $S(k)$, and hence the quantum DCF, is essentially 
the same as with pair correlations, the liquid energies are slightly
lowered.  This results in a consistent lowering also of the solid energies 
and a slight reduction in the densities of coexisting liquid and solid, 
indicating that comparisons with highly accurate simulation data, 
not restricted to pair correlations, should properly incorporate 
triplet correlations.
In comparison with more accurate GFMC simulation data~\cite{KLV}, however, 
the predicted solid energies and freezing densities are still significantly 
overestimated.
For clarity, it is important to distinguish between higher-order correlation
factors in the trial wavefunction of the liquid-state theory and higher-order 
DCFs in the approximate energy functional in the DF theory of the solid.  
Although here only the pair DCF is explicitly invoked through \ref{vr}, 
in principle the theory could be extended to include also higher-order DCFs, 
defined as higher-order functional derivatives of the energy functional,  
as in fact has been achieved already, at the level of the triplet DCF, 
in the classical theory~\cite{LA}.

The same procedure as above has been applied to the Fermi HS system, 
in this case though, using only the more accurate PPA-HNC+ input 
with pair correlations.  Figure~7 shows the Fermi liquid 
and {\it fcc} crystal energy densities together with the
Bose liquid energy and the corresponding VMC data of Schiff~\cite{Schiff} 
for comparison. 
We reiterate here that the Fermi liquid energy is approximated by the 
sum of the Bose liquid energy and the WF corrections shown in Figure~3.  
Of course, because the VMC data also are based on the WF expansion 
for the liquid and on the neglect of exchange effects in the solid, 
the comparison, although consistent, is not a test of the validity 
of these two approximations.
The predicted Fermi system freezing parameters are listed in Table~1, 
where the relative change of density upon freezing is seen to be about 4 \%.
Unfortunately, the scarce available simulation data~\cite{Schiff} do not 
permit an accurate determination of freezing parameters for comparison.
For convenience, the predicted energy densities for both Bose and Fermi 
systems from Figures~6~and~7 are listed numerically in Tables~2~and~3.
Note that the higher energy of the Fermi liquid compared with that 
of the Bose liquid simply reflects the destabilizing effect of 
Fermi statistics (exchange) on the liquid phase and naturally leads to 
a lowering of the liquid and solid coexistence densities, as is observed 
experimentally in the case of $^3$He and $^4$He~\cite{Wilks}.
Considering the sensitivity of the freezing parameters to 
details of the equation of state, however, quantitative predictions 
must await a more accurate treatment of the Fermi liquid~\cite{KD}.

The consistent over-prediction of solid energies points to a
systematic error either in the theory or in the input to the theory.
To investigate this issue, it is instructive to compare the present results 
with those found previously by a similar analysis~\cite{DNA}. 
In the previous study, empirical scaling of the liquid-state DCF 
by a factor of roughly 1.2 considerably improved the freezing parameters.  
Similar conclusions have been reached by other workers~\cite{MS1,Dalfovo} 
who applied the RY version of DF theory to $^4$He.
On the assumption that the MWDA is basically reliable, at least 
for HS systems, this observation would suggest that the true $v(k)$ 
is considerably more structured than that predicted by the current 
PPA-based liquid-state theory.  
Indeed, calculations and comparison with experiments for liquid $^4$He 
have shown~\cite{krot3,string} that the Feynman approximation tends to 
underestimate the structure of the static response function, especially 
in the crucial region near the roton maximum.
This is evidence that the incorrect asymptotic behaviour of $v^F(k)$
(see Appendix) may be adversely affecting the relevant finite-$k$ region. 
In quantum freezing, the $k$-range of practical interest is considerably 
narrower than in classical systems.  This is because the quantum solid 
has a much lower density near the transition than the classical solid, 
and hence a much wider Gaussian density distribution.  
Included in Figure~2 are the positions and relative weights of the 
first few {\it fcc} reciprocal lattice vectors, illustrating that the region 
of the first minimum, and to a lesser extent the second minimum, of $v(k)$
is crucial in determining the weighted density and thus the energy of 
the solid.  The fact that $S(k)$ is relatively insensitive to
the type of closure approximation used (HNC or HNC+) suggests that it is
the Feynman approximation [equation \ref{Feynmanv}], in converting $S(k)$ to 
$v(k)$, that is most likely responsible for any significant error in $v(k)$.  
Finally, it should be noted that any remaining high-density error in the 
PPA calculation of the input liquid energy will have negligible effect 
on the solid energy, since the MWDA always maps the solid onto 
an effective liquid of lower density (see Figure~5).
Thus, the only option for improving the theoretical results appears to lie 
in the input of a more accurate liquid-state DCF, respecting the correct
asymptotic limit.

\section{Summary and Conclusions}
By way of recapitulation, we have applied density-functional theory of 
non-uniform systems to quantum hard-sphere solids at zero temperature, and 
thereby studied the liquid-solid transition of Bose and Fermi systems.
Mapping both the Bose and Fermi solids onto the corresponding Bose liquid 
through the use of the modified weighted-density approximation, 
ignoring exchange effects in the Fermi solid, 
and taking liquid-state input data from an accurate paired phonon analysis
calculation coupled with the Feynman approximation, the solid energy and 
freezing parameters have been computed with no adjustable parameters.
The qualitative influence of Fermi statistics on the freezing transition
also has been considered by using the Wu-Feenberg cluster expansion method
to approximate the effect of antisymmetry on the Fermi liquid energy.
Compared with the Bose liquid, the energetically less stable Fermi liquid 
freezes at lower density into a lower-density solid 
with a higher Lindemann ratio.

The ground-state energies obtained using liquid-state input data of
varying degrees of accuracy are generally in satisfactory qualitative 
agreement with available simulation data.  Furthermore, distinct liquid-solid
transitions are always obtained, demonstrating a certain robustness 
of the theory.  
Quantitative accuracy depends heavily, however, on the accuracy
of the input data.  The freezing parameters, in particular, are 
especially sensitive to the detailed form of the equation of state.
As shown previously for Bose hard spheres~\cite{DNA}, empirical scaling 
of the Feynman DCF (or static response function) by an appropriate
structure-enhancing factor results in much closer agreement with 
simulation than is obtained by using the best available unscaled 
liquid-state data.
Experience with $^4$He further indicates that the Feynman approximation
tends to underestimate the structure of the liquid. 
Finally, analysis of the short-wavelength asymptotic behaviour of the 
exact liquid DCF reveals that the Feynman DCF tends to the wrong limit.
We conclude therefore that a more accurate liquid-state DCF -- 
generated either through alternatives to the Feynman approximation or 
by Quantum Monte Carlo simulation -- is still called for to achieve a 
decisive quantitative test of the theory in this demanding application.

\ack
We are grateful to E Krotscheck for his interest in the work and 
for supplying the PPA data, and to C N Likos for sending a preprint 
of Ref.~\cite{CL} prior to publication.
This work was supported by the Materials Science Center at
Cornell University and through NSF Grant No DMR 9319864.
ARD gratefully acknowledges support at various stages 
from the Natural Sciences and Engineering Research Council of Canada 
and from the Austrian Science Foundation.
PN thanks the DFG (Heisenberg Foundation) for support.

\Appendix{Asymptotic Limit of $v(k)$}

As discussed in Sec.~3, analysis of the lowest-order frequency moments of 
$S(k,\omega)$ leads directly to the Feynman approximation for $v(k)$, which 
is seen [from \ref{Feynmanv} and \ref{smallk}] to have the asymptotic limit
$v^F(k) \to 0$ as $k \to \infty$.  
The moment analysis may be straightforwardly extended~\cite{string}
to include the higher-order moments $m_2$ and $m_3$.  
The latter may be expressed in the form~\cite{HF,string}
$$m_2 = \Bigl({{\hbar^2k^2}\over{2m}}\Bigr)^2\Bigl[2-S(k)\Bigr] +
{{\hbar^4k^2}\over{m^2}} D(k),\label{m2-a}$$
where $D(k)$ is the kinetic structure function, defined by
$$\fl D(k) \equiv \int d{\bdi r}\int d{\bdi r}' 
\cos(k(z-z'))\nabla_z\nabla_{z'}
\rho^{(2)}({\bdi r},{\bdi r}';{\bdi R},{\bdi R}')|_{\{{\bdi r,r}'\}
=\{{\bdi R,R}'\}},\label{m2-b}$$
and
$$\fl m_3 = \Bigl({{\hbar^2k^2}\over{2m}}\Bigr)^3 + 
\Bigl({{\hbar^2k^2}\over{m}}\Bigr)^2 {\langle}\epsilon_K{\rangle} + 
{{\hbar^4\rho}\over{2m^2}}\int d{\bdi r} g(r)[1-\cos({\bdi k}\cdot {\bdi r})]
({\bdi k}\cdot\nabla)^2\phi(r),\label{m3}$$
with ${\langle}\epsilon_K{\rangle}$ the mean ground-state kinetic energy 
per particle and $\phi(r)$ the interatomic potential.
Starting from the rigorous inequality
$$\int_0^{\infty}d\omega {{S(k,\omega)}\over{\hbar\omega}}
(1+a\hbar\omega+b\hbar^2\omega^2)^2 \geq 0,\label{ineq2}$$
and minimizing the left-hand side with respect to the real parameters 
$a$ and $b$, yields the bound
$$m_{-1}(k) \geq {{m^F_{-1}(k)}\over{1-\Delta(k)/\epsilon(k)}},
\label{mbound2}$$
where
$$\Delta(k)={m_2\over{m_1}}-{m_1\over{m_0}}\label{Delta}$$
and
$$\epsilon(k)=\Bigl[{m_3\over{m_1}}+\Bigl({m_1\over{m_0}}\Bigr)^2-
2{m_2\over{m_o}}\Bigr]/\Delta(k).\label{epsilon}$$
Treating \ref{mbound2} as an equality, and using \ref{vchi}, \ref{chio2}, and 
\ref{m-a}, leads to the approximation
$$v(k) = {{\hbar^2k^2}\over{4m\rho}}\Bigl[{1\over{S^2(k)}}
\Bigl(1-\Delta(k)/\epsilon(k)\Bigr)-1\Bigr],\label{String}$$
which in principle represents an improvement over the Feynman approximation.  
Indeed, for $^4$He the corresponding density-density 
static response has been explicitly computed by quantum MC and 
shown to significantly improve the comparison with experiment~\cite{CS1}.
Since the kinetic structure function is not readily available for hard 
spheres, however, we have not attempted to use \ref{String} in this paper.  
Nevertheless, it is interesting to consider the asymptotic short-wavelength
limit.  This is straightforward, since it is known~\cite{HF} that
$$D(k) \to {2\over{3}}{m\over{\hbar^2}}\langle\epsilon_K\rangle
\quad (k \to \infty)\label{D}$$
and since for hard spheres $m_3(k)$ may be explicitly expressed as~\cite{Puff}
$$\fl m_3 = \Bigl({{\hbar^4k^4}\over{2m^2}}\Bigr)mc^2 + 
\Bigl({{\hbar^2k^2}\over{2m}}\Bigr)^3 + \Bigl({{\hbar^2k^2}\over{m}}\Bigr)^2
\Bigl[\Bigl({P\over{\rho}}-{2\over{3}}{\langle}\epsilon_K\rangle\Bigr)
\Bigl({\cal P}_1(k\sigma)-1\Bigr)\cr
\lo+ \Bigl(mc^2-{{2P}\over{\rho}}-{2\over{3}}
{\langle}\epsilon_K\rangle\Bigr)\Bigl({\cal P}_2(k\sigma)-
{1\over{2}}\Bigr)\Bigr],\label{m3hs-a}$$
where $P$ is the pressure and
$${\cal P}_1(x) \equiv {3\over{x^2}}\Bigl({{{\sin}x}\over{x}}-{\cos}x\Bigr),
\label{m3hs-b}$$
$${\cal P}_2(x) \equiv {5\over{x^2}}\Bigl[{1\over{3}}+{2\over{x^2}}
\Bigl({{{\sin}x}\over{x}}-{\cos}x\Bigr)-{{{\sin}x}\over{x}}\Bigr].
\label{m3hs-c}$$
The various moments appearing in \ref{Delta} and \ref{epsilon} are easily
shown to behave asymptotically as
$$m_0 \to 1,\label{asym-a}$$
$$m_1 \to {{\hbar^2k^2}\over{2m}},\label{asym-b}$$
$$m_2 \to \Bigl({{\hbar^2k^2}\over{2m}}\Bigr)^2+{2\over{3}}
{{\hbar^2k^2}\over{m}}\langle\epsilon_K\rangle,\label{asym-c}$$
and
$$m_3 \to \Bigl({{\hbar^2k^2}\over{2m}}\Bigr)^3+\Bigl({{\hbar^2k^2}\over{m}}
\Bigr)^2\langle\epsilon_K\rangle.\label{asym-d}$$
Substituting these limits into \ref{Delta} and \ref{epsilon}, we obtain
$$\Delta(k)/\epsilon(k) \to {8\over{3}}{{m\langle\epsilon_K\rangle}
\over{\hbar^2k^2}},\label{ratio}$$
which, upon further substitution into \ref{String}, yields
$$v(k) \to -{2\over{3}}{{\langle\epsilon_K\rangle}\over{\rho}}.\label{vslim}$$
Thus, the exact $v(k)$ approaches a {\it finite} negative constant
as $k \to \infty$, in sharp contrast to the asymptotic vanishing of the
Feyman approximation, as well as of the classical DCF.  It is interesting 
to note that analogous behaviour has been derived for the corresponding 
function defined for the uniform electron liquid~\cite{Holas}.  
In fact, as has been recently suggested~\cite{CL}, it is quite probably 
a universal feature of quantum DCFs.
What this curious asymptotic behaviour may imply for the {\it finite}-$k$ 
behaviour of the true $v(k)$ of Bose hard spheres remains to be clarified 
by a decisive quantum MC computation. 

\vfil\eject
\numreferences

\bibitem{DF1} 
For reviews of the basic principles of DF theory see 
Evans R 1979 {\it Adv. Phys.} {\bf 28} 143;
Evans R 1989, in {\it Liquids at Interfaces}, Les Houches session 48, edited
by Charvolin J, Joanny J F and Zinn-Justin J (New York: Elsevier);
Oxtoby D W, 1990 in {\it Liquids, Freezing and Glass Transition}, Les
Houches session 51, edited by Hansen J-P, Levesque D and Zinn-Justin J
(New York: Elsevier)

\bibitem{RY} 
Ramakrishnan T V and Yussouff M 1979 {\it Phys. Rev.} B {\bf 19} 2775;
Haymet A D J and Oxtoby D W 1981 {\it J. Chem. Phys.} {\bf 74} 2559

\bibitem{HM} 
Hansen J-P and McDonald I R 1986 {\it Theory of Simple Liquids}
$2 ^{nd}$ edition (London: Academic)

\bibitem{DF2} 
For recent reviews of applications of DF theory see 
L\"owen H {\it Phys. Reports} 1994 {\bf 237} 249;
Singh Y 1991 {\it Phys. Reports} {\bf 207} 351

\bibitem{QDF} 
For reviews of quantum DF theory see Moroni S and Senatore G 1994 
{\it Phil. Mag.} B 
{\bf 69} 957; Haymet A D J 1992 in {\it Inhomogeneous Fluids} edited by 
Henderson D (New York: Dekker)

\bibitem{MS1} 
Moroni S and Senatore G 1991 {\it Europhys. Lett.} {\bf 16} 373

\bibitem{Dalfovo} 
Dalfovo F, Dupont-Roc J, Pavloff N, Stringari S and Treiner J 1991
{\it Europhys. Lett.} {\bf 16} 205 

\bibitem{SP} 
Senatore G and Pastore G 1990 {\it Phys. Rev. Lett.} {\bf 64} 303

\bibitem{MS2} 
Moroni S and Senatore G 1991 {\it Phys. Rev.} B {\bf 44} 9864

\bibitem{CG} 
Choudhury N and Ghosh S K 1995 {\it Phys. Rev.} B 51 2588

\bibitem{DNA} 
Denton A R, Nielaba P, Runge K J and Ashcroft N W 1990 {\it Phys. Rev. Lett.}
{\bf 64} 1529; 1991 {\it J. Phys. Condens. Matter} {\bf 3} 593

\bibitem{MRH} 
McCoy J D, Rick S W and Haymet A D J 1989 {\it J. Chem. Phys.} {\bf 90}
4622; 1990 {\it J. Chem. Phys.} {\bf 92} 3034

\bibitem{RMH} 
Rick S W, McCoy J D and Haymet A D J 1990 {\it J. Chem. Phys.}
{\bf 92} 3040

\bibitem{CS1} 
Moroni S, Ceperley D M and Senatore G 1992 {\it Phys. Rev. Lett.} {\bf 69} 1837

\bibitem{CS2} 
Moroni S, Ceperley D M and Senatore G 1995 {\it Phys. Rev. Lett.} {\bf 75} 689

\bibitem{MWDA} 
Denton A R and Ashcroft N W 1989 {\it Phys. Rev.} A {\bf 39} 4701

\bibitem{PPA} 
Chang C C and Campbell C E 1977 {\it Phys. Rev.} B {\bf 15} 4238;
Jackson H W and Feenberg E 1961 {\it Ann. Phys.} {\bf 15} 266

\bibitem{KS} 
Krotscheck E and Saarela M 1993 {\it Phys. Rep.} {\bf 232} 1;
Krotscheck E 1986 {\it Phys. Rev.} B {\bf 33} 3158

\bibitem{WF} 
Wu F Y and Feenberg E 1962 {\it Phys. Rev.} {\bf 128} 943;
Pandharipande V R and Bethe H A 1973 {\it Phys. Rev.} C {\bf 7} 1312

\bibitem{HK} 
Hohenberg P and Kohn W 1964 {\it Phys. Rev.} {\bf 136} B864

\bibitem{WDA} 
Curtin W A and Ashcroft N W 1985 {\it Phys. Rev.} A {\bf 32} 2909;
1986 {\it Phys. Rev. Lett.} {\bf 56} 2775; 1986 {\bf 57} 1192 erratum

\bibitem{HLS} 
Hansen J-P and Levesque D 1968 {\it Phys. Rev.} {\bf 165} 293;
Hansen J-P, Levesque D and Schiff D 1971 {\it Phys. Rev.} A {\bf 3} 776

\bibitem{CCK} 
Ceperley D, Chester, G V and Kalos M H 1977 {\it Phys. Rev.} B {\bf 16} 3081

\bibitem{McM} 
McMillan W L 1965 {\it Phys. Rev.} {\bf 138} A 442

\bibitem{HF} 
Hall D and Feenberg E 1971 {\it Ann. Phys.} {\bf 63} 335

\bibitem{CR} 
Chester G V and Reatto L 1966 {\it Phys. Lett.} {\bf 22} 276

\bibitem{CFK} 
Campbell C E, Folk R and Krotscheck E 1997 {\it J. Low Temp. Phys.}
in press

\bibitem{CL} 
Likos C N, Moroni S and Senatore G 1997 {\it Phys. Rev.} B in press

\bibitem{Schiff} 
Schiff D 1973 {\it Nature Phys. Sci.} {\bf 243} 130

\bibitem{Brandow} 
Brandow B H 1976 {\it Phys. Lett.} B {\bf 61} 117

\bibitem{krot1} 
Krotscheck E 1976 {\it Lett. Nuov.} C {\bf 16} 269;
Krotscheck E and Takahashi K 1976 {\it Phys. Lett.} B {\bf 63} 269

\bibitem{krot2} 
Krotscheck E 1977 {\it Nucl. Phys.} A {\bf 293} 293;
1977 {\it J. Low Temp. Phys.} {\bf 27} 199

\bibitem{KD} 
Krotscheck E and Denton A R unpublished

\bibitem{Wilks} 
Wilks J 1967 {\it The Properties of Liquid and Solid Helium} 
(Clarendon: Oxford University Press)

\bibitem{YA} 
Young D A and Alder B J 1974 {\it J. Chem. Phys.} {\bf 60} 1254

\bibitem{OLW} 
Ohnesorge R, L\"owen H and Wagner H 1993 {\it Europhys. Lett.}
{\bf 22} 245

\bibitem{WCCK} 
Whitlock P A, Ceperley D M, Chester G V and Kalos M H 1979
{\it Phys. Rev.} B {\bf 19} 5598

\bibitem{KLV} 
Kalos M H, Levesque D and Verlet L 1974
{\it Phys. Rev.} A {\bf 9} 2178

\bibitem{LA} 
Likos C N and Aschcroft N W 1992 {\it Phys. Rev. Lett.} {\bf 69} 316;
1992 {\bf 69} 3141E; 1993 {\it J. Chem. Phys.} {\bf 99} 9090

\bibitem{krot3} 
Krotscheck E private communication.

\bibitem{string} 
Stringari S 1992 {\it Phys. Rev.} B {\bf 46} 2974; 
Dalfovo F and Stringari S 1992 {\it Phys. Rev.} B {\bf 46} 13991; 
1992 {\it J. Low Temp. Phys.} {\bf 89} 325

\bibitem{Puff} 
Puff R D 1965 {\it Phys. Rev.} {\bf 137} A 406

\bibitem{Holas} 
Holas A 1987 in {\it Strongly Coupled Plasma Physics} 
edited by Rogers F J and Dewitt H E (New York: Plenum)

\Tables

\table
{Freezing parameters for the ground-state Bose and Fermi hard-sphere systems:
Liquid and solid ({\it fcc}-crystal) coexistence densities, $\rho_l$ and 
$\rho_s$, density change $\Delta\rho$, Gaussian width parameter 
$\alpha$, and Lindemann ratio $L$.}
\align\L{#}&&\C{#}\cr
\br
 &$\rho_l\sigma^3$&$\rho_s\sigma^3$&$\Delta\rho\sigma^3$
&$\alpha\sigma^2$&$L$\cr
\br
{\bf Bose Hard Spheres}:& & & & & \cr
Theory (PPA-HNC)&0.351&0.365&0.014&14.38&0.206\cr
Theory (PPA-HNC+)&0.359&0.371&0.012&13.39&0.214\cr
Simulation&0.23$\pm$.02&0.25$\pm$.02&0.02&6.25&0.27\cr
\mr
{\bf Fermi Hard Spheres}:& & & & & \cr
Theory (PPA-HNC+)&0.313&0.326&0.013&10.24&0.235\cr
\br
\endalign
\endtable

\table
{Liquid-state numerical data corresponding to Figures~6~and~7:
Predicted ground-state energy densities over a range of number densities
$\rho$ for the Bose hard-sphere liquid ($E^B/V$) in the HNC and HNC+ 
approximations and for the Fermi hard-sphere liquid ($E^F/V$) in the HNC+ 
approximation.}
\align\C{#}&&\C{#}\cr
\br
$\rho\sigma^3$&$E^B/V$ (HNC)&$E^B/V$ (HNC+)&$E^F/V$ (HNC+)\cr
\br
0.02& 0.0043& 0.0044& 0.0068\cr
0.04& 0.0205& 0.0204& 0.0290\cr
0.06& 0.0539& 0.0524& 0.0702\cr
0.08& 0.1099& 0.1052& 0.1345\cr
0.10& 0.1945& 0.1842& 0.2266\cr
0.12& 0.3141& 0.2954& 0.3521\cr
0.14& 0.4760& 0.4452& 0.5174\cr
0.16& 0.6883& 0.6411& 0.7295\cr
0.18& 0.9601& 0.8909& 0.9964\cr
0.20& 1.3018& 1.2036& 1.3271\cr
0.22& 1.7252& 1.5889& 1.7313\cr
0.24& 2.2434& 2.0574& 2.2198\cr
0.26& 2.8713& 2.6210& 2.8045\cr
0.28& 3.6257& 3.2925& 3.4981\cr
0.30& 4.5252& 4.0857& 4.3147\cr
0.32& 5.5906& 5.0159& 5.2693\cr
0.34& 6.8451& 6.0996& 6.3781\cr
0.36& 8.3141& 7.3546& 7.6586\cr
0.38&10.0257& 8.8000& 9.1294\cr
\br
\endalign
\endtable

\table
{Solid-state numerical data corresponding to Figures~6~and~7:
Predicted ground-state energy densities over a range of number densities
$\rho$ for the hard-sphere solid ($E/V$) -- independent of statistics --
in the HNC and HNC+ approximations.}
\align\C{#}&&\C{#}\cr
\br
$\rho\sigma^3$&$E/V$ (HNC)&$E/V$ (HNC+)\cr
\br
0.26& 3.3346& 3.0779\cr
0.28& 4.0648& 3.7279\cr
0.30& 4.8955& 4.4608\cr
0.32& 5.8279& 5.2754\cr
0.34& 6.9521& 6.2449\cr
0.36& 8.2973& 7.3914\cr
0.38& 9.8544& 8.7044\cr
0.40&11.6209&10.1785\cr 
0.42&13.5754&11.8113\cr
0.44&15.7220&13.5863\cr
\br
\endalign
\endtable

\Figures

\figure
{Static structure factor $S(k)$ vs. wave vector magnitude $k$ 
for Bose hard-sphere liquids of three different densities, generated by 
the PPA with the HNC+ approximation (see text).} 

\figure
{Quantum direct correlation function $v(k)$, corresponding to 
$S(k)$ in Figure~1, computed from the Feynman approximation
[equation (26)].  Vertical bars indicate
positions of {\it fcc} reciprocal lattice vectors at average solid density
$\rho_s\sigma^3=0.4$, for which the Gaussian width parameter 
$\alpha\sigma^2=15.8$.
Bar heights, arbitrarily normalized to 50, are proportional 
to the product of degeneracy and the Gaussian factor in (34), and
thus represent relative contributions to the weighted density $\hat\rho$.}

\figure
{First three terms in the Wu-Feenberg cluster expansion, representing
Fermi-statistics corrections to the Bose hard-sphere liquid energy 
per particle (units of $\hbar^2/m\sigma^2$).}

\figure
{Hard-sphere solid energy density (units of $\hbar^2/m\sigma^5$) 
vs. Gaussian width parameter (solid line) for average solid density 
$\rho_s\sigma^3=0.37$.  Shown separately are ideal-gas energy 
(short-dashed line) and correlation energy (long-dashed line).}

\figure
{Weighted density vs. average solid density at the energy minimum
for the hard-sphere {\it fcc} crystal.}

\figure
{Bose hard-sphere liquid and solid energy densities vs. density,  
generated for the liquid by the PPA and for the solid by 
density-functional theory (DFT).  Dashed lines correspond to the simple 
HNC approximation, solid lines to the enhanced HNC+ approximation (see text).  
Circles and squares denote variational Monte Carlo (VMC) data~\cite{HLS} 
for the liquid and solid, respectively.}

\figure
{Fermi hard-sphere liquid and solid energy densities vs. density, 
generated for the liquid by the PPA, with the HNC+ approximation and
Wu-Feenberg corrections (see Figure~3 and text), and for the solid by DF 
theory.  Dashed line shows the Bose liquid energy density from Figure~6,
illustrating the approximate magnitude of statistics corrections in the liquid.
Circles and squares denote VMC data~\cite{Schiff}
for the liquid and solid, respectively.}

\bye